# Towards using Large Language Models for Risk Negotiation and Consensus-building in One Health contexts

Alexandra Fetsch [1†], Iurii Savvateev[2†], Racem Ben Romdhane[2], Martin Wiedmann[3], Artemiy Dimov[4,5,6], Maciej Durkalec[7,8], Josef Teichmann[9], Jakob Zinsstag[4,5], Konstantinos Koutsoumanis[10], Andreja Rajkovic[11], Jason Mann[12], Mauro Tonolla[13], Monika Ehling-Schulz[14], Matthias Filter[2], Sophia Johler[1*]

[1] Chair of Food Microbiology, Faculty of Veterinary Medicine, LMU Munich, Munich, Germany; Email: alexandra.fetsch@lmu.de, sophia.johler@lmu.de

[2] Study Centre Supply Chain Modelling and Artificial Intelligence, German Federal Institute for Risk Assessment, Berlin, Germany; Email: Iurii.Savvateev@bfr.bund.de, racem.ben-romdhane@bfr.bund.de, matthias.filter@bfr.bund.de

[3] Department of Food Science, Cornell University, Ithaca, USA; E-mail: martin.wiedmann@cornell.edu

[4] Swiss Tropical and Public Health Institute, Allschwil, Switzerland; Email: artemiy.dimov@swisstph.ch, jakob.zinsstag@swisstph.ch

[5] University of Basel, Petersplatz 1, 4001 Basel, Switzerland

[6] Institute for Food Safety and Hygiene, Vetsuisse Faculty, University of Zurich, Zurich, Switzerland

[7] Study Centre for Land- use related Evaluation procedures, One-Health, German Federal Institute for Risk Assessment, Berlin, Germany; Email: Maciej.Durkalec@bfr.bund.de

[8] Department of Pharmacology and Toxicology, National Veterinary Research Insitute, Puławy, Poland

[9] Department of Mathematics, ETH Zurich, Zurich, Switzerland; Email: jteichma@math.ethz.ch

[10] Department of Food Science and Technology, Aristotle University of Thessaloniki, Thessaloniki, Greece; E-mail: doulgeraki@agro.auth.gr

[11] Faculty of Bioscience Engineering, Department. of Food Technology, Safety and Health, Ghent University, Ghent, Belgium**,** E-mail: Andreja.Rajkovic@UGent.be
[12] Nestlé USA, Inc., Arlington, USA; Email: jason.mann@us.nestle.com
[13] Institute of Microbiology, University of Applied Sciences and Arts of Southern Switzerland, Mendrisio, Switzerland, Email: mauro.tonolla@supsi.ch
[14] Institute for Microbiology, Center of Pathobiology, Department of Biological Sciences and Pathobiology, University of Veterinary Medicine Vienna, Vienna, Austria; E-mail: Monika.Ehling-Schulz@vetmeduni.ac.at

† contributed equally

***Corresponding author:** Prof. Dr. Sophia Johler
**Phone** +49 89 2180 78 600
**Fax** +49 89 2180 78 602
**Email** sophia.johler@lmu.de

## Graphical Abstract

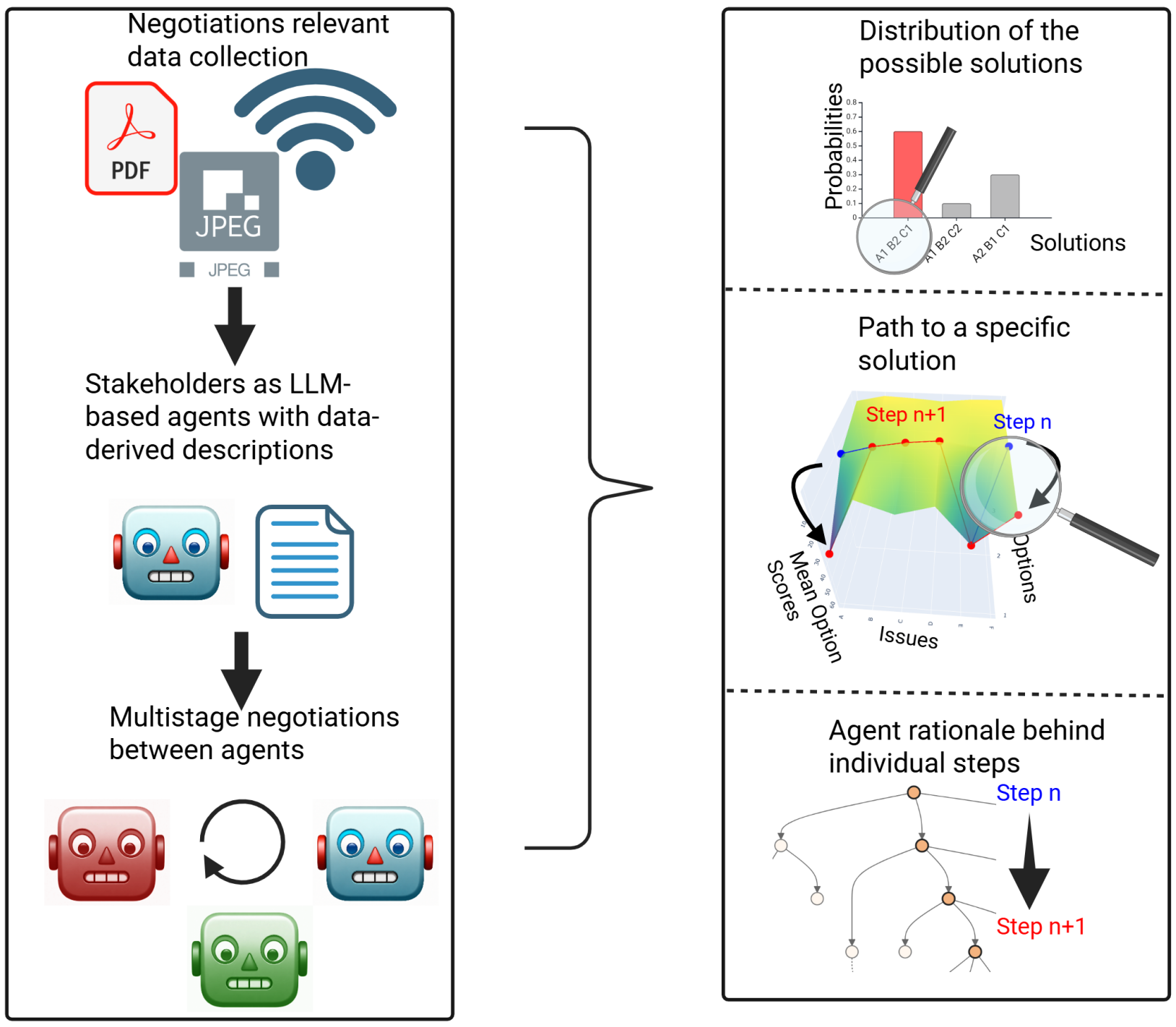

**Summary**

Risk negotiation under the One Health framework requires balancing competing stakeholder interests across human, animal, and environmental health domains — a process constrained by information complexity and time pressure. This study presents and evaluates a multi-step pipeline that integrates large language model-based agents into a human-supervised negotiation process, specifically targeting risk assessment, valuation, and negotiation steps. Agent descriptions were generated from stakeholder-provided materials using retrieval-augmented generation methods. During preliminary multi-agent discussions, agents accessed domain-specific computational models and web-based sources before human participants refined proposed issues and options. A simulated negotiation game then produced distributions of candidate deals, characterized using Nash bargaining solution criteria and Pareto optimality metrics. The developed approach was evaluated in a workshop involving independent domain experts across two real-world scenarios: food safety criteria for Campylobacter in broiler meat and oak processionary moth management. In this feasibility evaluation, participants reached consensus with the allotted time after discussing and modifying the proposed deals. The framework offers an adaptable blueprint for AI-assisted risk negotiation; the specific contribution of LLM assistance, including gains in efficiency and decision quality, requires comparative evaluation.

## Introduction

The increasing frequency and severity of threats such as infectious disease pandemics highlight the interconnectedness of ecosystems, animal health, and human health, and the need for proactive, preventative approaches to mitigate these risks [1,2]. To address these multidimensional challenges, integrative frameworks such as One Health have gained growing recognition [3], emphasizing the links between the health of humans, animals, plants, and the environment, and calling for integrated, participatory efforts [4,5]. Decision-making in this context is inherently complex, requiring the consideration of diverse parties' (i.e. stakeholders') interests and balancing specific and overall public health concerns with economic and social factors [6]. One key challenge is the area of risk analysis, which involves identifying potential threats, assessing their severity and likelihood, and determining the most effective course of action [7-9]. Particularly in the field of food safety, experts call for a move beyond binary compliance toward an integrated, risk-based approach that transparently defines appropriate levels of protection and balances safety with sustainability, equity, and trade objectives to enable resilient and secure food systems transformation[10].

Recently, we introduced the concept of a negotiation-centered framework for multi-sector risk analysis in the Bulletin of the World Health Organization [9]. In our concept, negotiation is understood in line with the definition of negotiations as discussions "between parties with perceived divergence of interests" that aim "to reach agreement on the distribution of scarce resources, work procedures, the interpretation of facts, or some commonly held opinion or belief" [8]. In this context, negotiation supports sustainable and efficient management of risks [8]. Practically, the negotiation process incorporates the following key steps: (i) selection of stakeholder representatives and formation of the roundtable; (ii) problem formulation by the roundtable parties; (iii) risk assessment and valuation, i.e. assessing the risks associated with various action options aimed at addressing the problem; (iv) risk negotiation among the stakeholders to agree on balanced action options; (v) communicating the negotiation outcomes to the respective stakeholder groups, which then proceed with implementing the agreed actions; and (vi) evaluation of the outcomes of the actions implemented and, if necessary, initiating a re-negotiation process.

The current work aimed to develop, test and evaluate a practical approach to augment this negotiation-centered risk analysis, particularly steps (iii) and (iv), with LLM-based multi-agent modelling[9] under human supervision (Figure 1). The term "multi-agent simulation" refers to an orchestrated interaction among multiple role-conditioned LLM instances representing distinct stakeholders. The agents generate responses and may invoke predefined tools within individual turns. Rather than testing whether LLM assistance outperforms unaided human negotiation, we investigated whether domain-specific knowledge (step iii) and conflicting stakeholder preferences (step iv) can be integrated within a structured computational workflow that generates explicit issue-option formulations and candidate compromise deals, which are then evaluated by human participants. To this end, first, a customized retrieval-augmented generation (RAG) approach [11] and few-shot LLM calls [12] were applied to generate descriptions from multimodal materials-including text files, images, and relevant online sources-provided by the corresponding human participants. The descriptions were then used to construct custom prompts for LLM-based agents participating in a preliminary multi-agent discussion simulation, aimed at identifying topic-related issues and acceptable solution options. The agents were additionally equipped with tools enabling web search and API-based communication with domain-relevant models deposited as FSK-ML compliant models [13]. Following this data-driven issue and option identification, human participants specified the relative importance of each issue and their preferences among the solution options, thereby generating stakeholder preference scores. These scores, together within the issue–option descriptions, were used to simulate a non-zero-sum negotiation game within the LLM-based multi-agent framework [14,15]. The feasibility of the practical implementation was evaluated by independent field practitioners in two real-world risk negotiation scenarios: (i) setting the food safety criterion for *Campylobacter* spp. in broiler meat; and (ii) management of the oak processionary moth (*Thaumetopoea processionea*) in Central Europe.

We present a proof-of-concept for a practical tool intended to facilitate negotiation-centered risk management [9] across disciplines and real-world challenges. If confirmed in comparative evaluations, such implementations could accelerate the formulation of problem definitions, support consensus-building on risk prioritization and action selection, and facilitate the development of strategies to address identified risks in alignment with the 17 Sustainable Development Goals (SDGs) outlined in the UN 2030 Agenda [16]. More broadly, approaches of this kind have the potential to foster the dialogue between science and policymakers needed to translate political commitment into concrete action.

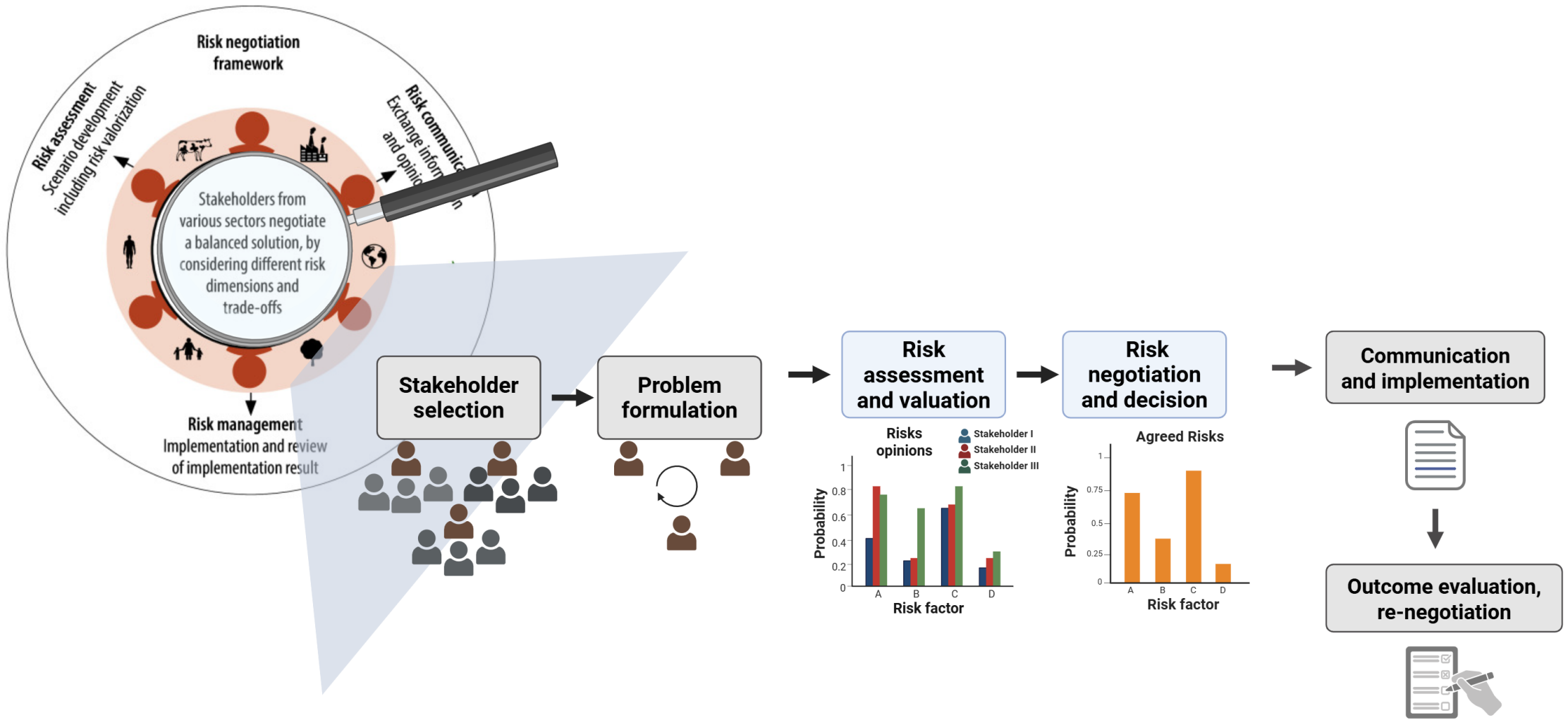

**Figure 1. Workflow of negotiation-centered risk analysis (incorporating the visualization of the risk negotiation framework from Ehling-Schulz *et al.* [9].** The figure visualizes the negotiation-centered risk analysis framework incorporating 6 different steps as follows (i) stakeholder round table establishment; (ii) problem formulation; (iii) risk assessment and valuation; (iv) risk negotiation; (v) communication and implementation; and (vi) outcome evaluation and optional risk re-negotiation. Blue highlighting indicates

the steps specifically addressed and further developed within the current work (steps iii and iv); remaining steps are shown in grey for contextual completeness.

## Results

In the present work, we (i) developed and tested a multi-step pipeline that integrates LLM-based agents into a human-in-the-loop (HIL) approach, and (ii) evaluated its feasibility through its application by external domain experts and stakeholders in a dedicated workshop setting focusing on two real-world case scenarios. Importantly, the proposed pipeline, the developed software solution, and the associated standard operating procedures (SOPs) (e.g., moderation protocol and stakeholder preparatory work) do not aim to be an ultimate solution but is designed to serve as a blueprint for integrating AI technologies into current risk negotiation processes.

The general workflow of our approach, and divided into steps 1-7, is schematically shown in Figure 2.

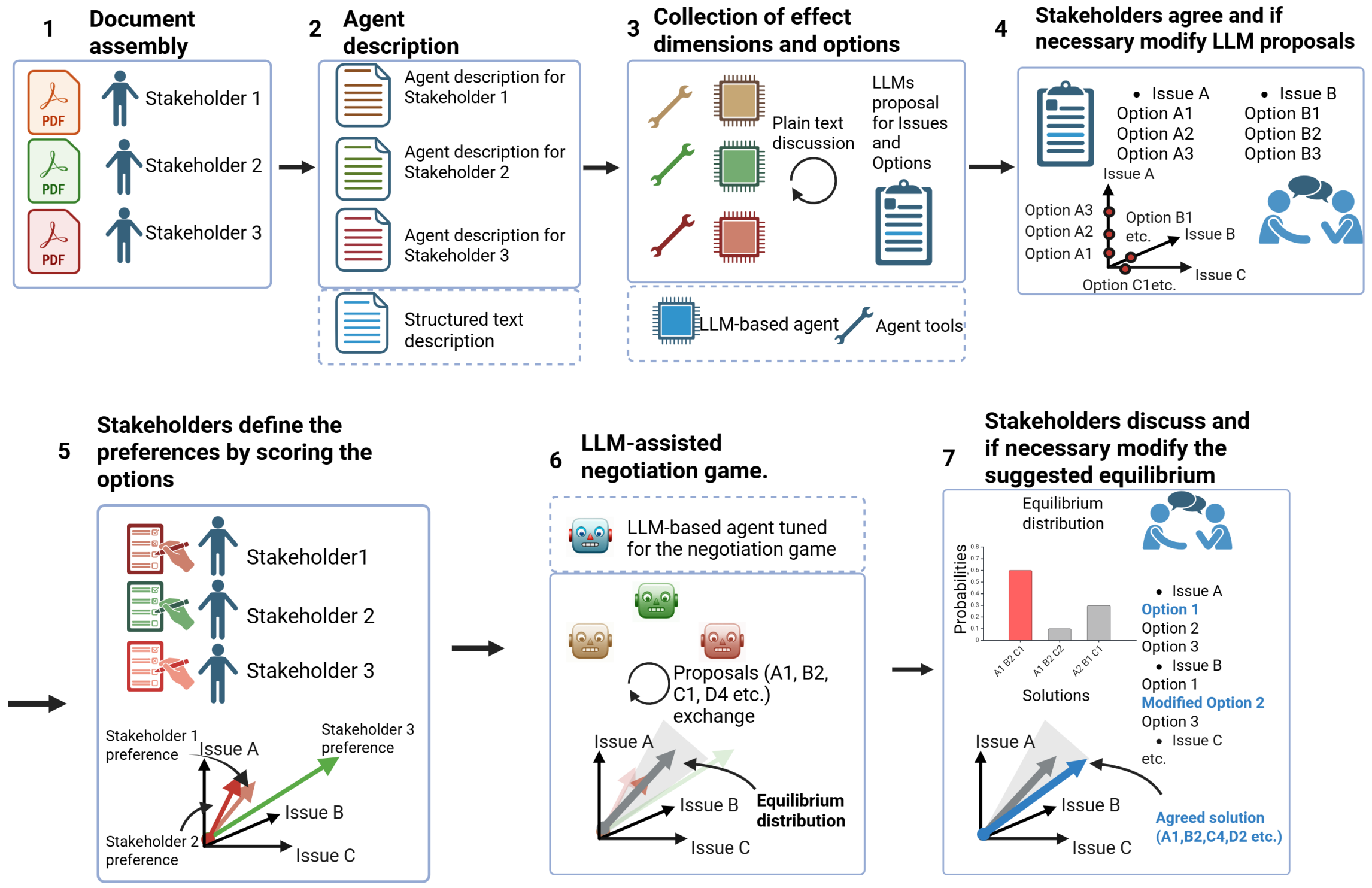

**Figure 2. Practical implementation of AI-assisted negotiation-centered risk analysis.** The figure illustrates the key steps from Figure 1 and how LLMs augmented the process. First, agent descriptions are generated from preselected literature and other source materials using RAG-based methods and few-shot LLM calls (*step 1*). These descriptions are then used to configure LLM-based agents and equip them with tools for the preliminary multi-agent discussion (*step 2*), aimed at breaking a broadly formulated problem down into discrete issues and possible solution options for each (*step 3*). Once stakeholders have agreed on issues and options to be negotiated (*step 4*), each stakeholder's stance is represented as a finite collection of scores, where each score corresponds to the value they assigned to a specific option within the high-dimensional space of issues (*step 5*). The negotiation process can then be visualized as a search for an equilibrium, where each stakeholder's scores act as attractors drawing the negotiation path toward their preferred outcomes. The resulting equilibrium point reflects a balance across these preferences (*step 6*). Importantly, the human-in-the-loop approach may cause the final approved deal to deviate from the simulated equilibrium, reflecting possible differences between real-world judgment and simulated agentic behavior (*step 7*).

In the following, we showcase the key elements of our approach-preliminary discussion, and main negotiations-using results from the two real-world case scenarios, targeting two different One Health problems during a workshop: (i) *Campylobacter* spp. in broiler meat, and (ii) *Thaumetopoea processionea* (oak processionary moth) in Central Europe, and involving several competing stakeholder interests (see Methods details for further characteristics, description of trade-offs etc.). Further, we outline the characterization of the LLM-suggested equilibrium consensus.

**Preliminary discussion: LLM-based agents are used for specifying issue-option groups**

Before entering the negotiation process, stakeholder representatives need to translate a broadly formulated problem into specific issues and identify possible action solutions (i.e., options) for addressing them (Figure 2, Steps 1-4). The process begins with the collection of documents and sources relevant to the general problem statement and containing information on prospective stakeholders, including their positions, key arguments, and previous lines of reasoning (Figure 2, Step 1). In the two studied case scenarios, these documents were provided by the invited participants or retrieved through classical literate evidence search before the negotiation exercises and consisted mainly of peer-reviewed scientific papers (see Zenodo for the sources used). The first case scenario examined how best to balance trade-offs for setting a Food Safety Criterion for *Campylobacter* spp. in broiler meat under EU legislation 2073/2005, with participants taking the roles of food safety authorities, food industry, environmental NGOs, and consumers. The second addressed the management of the oak processionary moth (OPM) in Central Europe, simulating competing environmental, public health, and land management interests. Below, we describe the preliminary negotiation using the *Campylobacter* spp. case scenario as an example, while the OPM case is used to demonstrate the equilibrium analysis of the main negotiation.

The collected documents for the *Campylobacter spp.* case scenario were used as input for RAG-based methods [17] to formulate the agent descriptions (Figure 2, Step 2). Specifically, we used two independent approaches to generate these descriptions. The first was LightRAG [11], an open-source approach that integrates knowledge graphs and vector representations, enabling improved information retrieval compared with isolated vector-based or graph-based RAG designs. The second was NotebookLM [18], a commercially available solution that uses a proprietary algorithm, presumably based on a RAG design, to ground responses in the provided materials and has recently been used in educational [19] and medical contexts [20]. The final agent descriptions were generated using OpenAI Deep Research, an LLM-based agentic system designed for multi-step research tasks [21]. Deep Research was prompted to compare the descriptions generated by LightRAG and NotebookLM, highlight their differences, and merge them into a final description. This procedure was intended to account for potential limitations inherent to either NotebookLM or LightRAG alone. For OPM case only NotebookLM followed by OpenAI Deep Research was used. The adequacy of the resulting descriptions was evaluated through an anonymous questionnaire administered at the end of the workshop; most participants (17 out of 18) expressed satisfaction with the RAG-based agent descriptions across both case scenarios, suggesting that the combined pipeline produced stakeholder representations that participants considered credible (see link to survey results in the key resources table).

Figure 3 schematically highlights the details of the preliminary discussion, once final agent descriptions were used to prompt LLM-based (i.e. GPT4o-based [22]) agents, each virtually representing a stakeholder. First, agents receive the current shared conversation history (Figure 3A, step 1), allowing them to perform a role-based analysis (Figure 3A, step 2). To do so, the agents were equipped with two groups of tools, in addition to LLM-powered capabilities of text analysis and generation [12,23]. The first enabled web-based information retrieval via arxiv.org [24], wikipedia.org [25], and duckduckgo.com [26]. The second linked LLM-based agents with web-hosted microbiological computational models from the RAKIP model repository [13,27] , enabling the use of domain-specific simulation results during the preliminary negotiation round (Figure 3A, step 3). Specifically, in the *Campylobacter* spp. case scenario, the agents called two computational models: one evaluating the impact of implementing a process hygiene criterion (PHC) for *Campylobacter* on broiler meat [28], and another assessing the effectiveness of three-parameter EU sampling plans for *Campylobacter* in poultry [29]. The agents engaged in a multi-round, plain-text conversation through a shared conversation history, with each agent analyzing the common history from the perspective of their assigned role as specified in the RAG-derived prompt and enriching their responses with tool-based evidence where appropriate (Figure 3A, steps 4-6). The conversation records were subsequently used in a few-shot GPT-4o call to generate a list of suggested issues in a format compatible with the subsequent negotiation step [14] - see Figure 3B, 3C for examples.

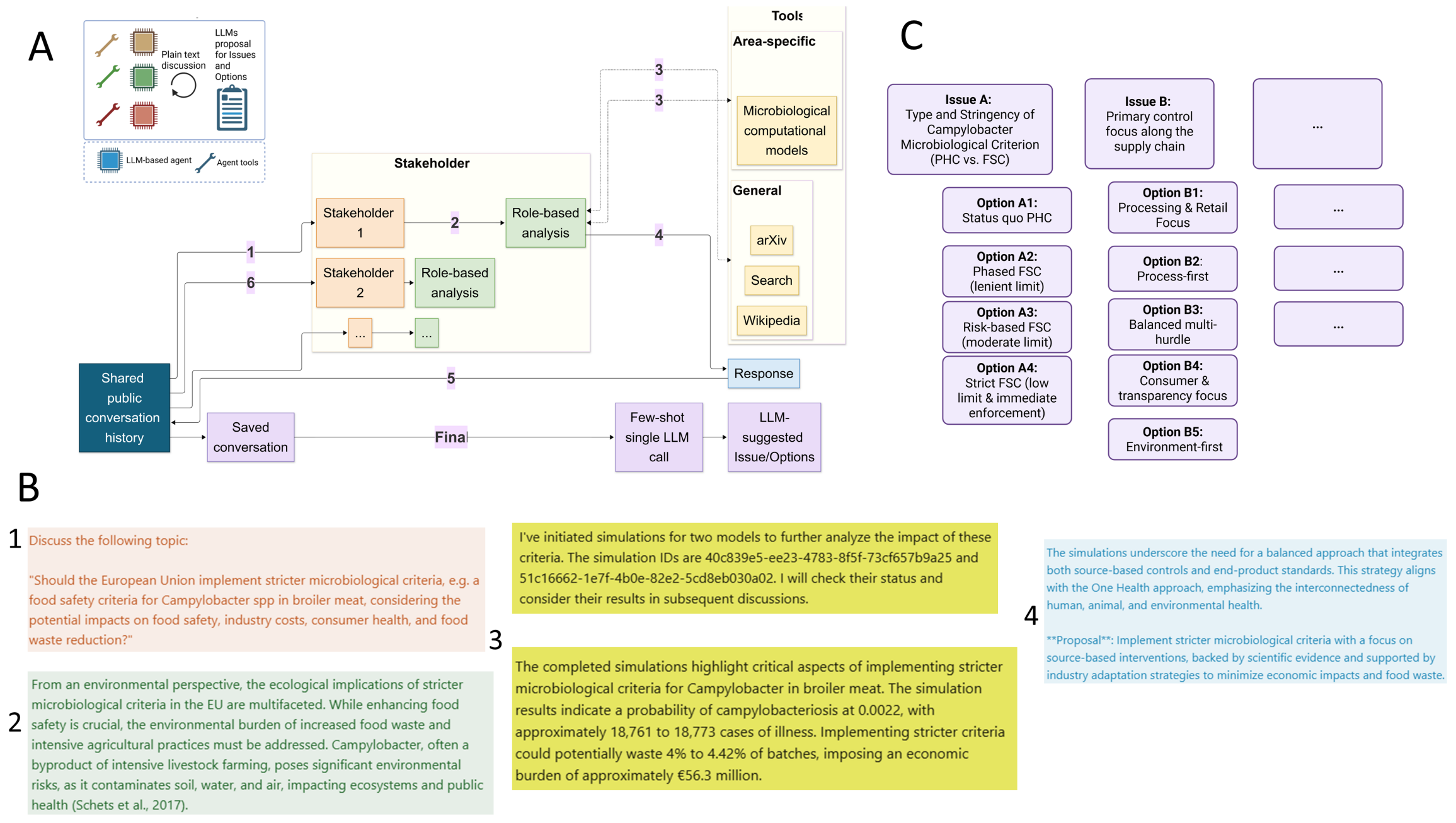

**Figure 3. Preliminary Discussion**. **(A)** Schema of the preliminary discussion, illustrating key steps (1-5) performed iteratively for each stakeholder. *Step 1*: the stakeholder receives the current shared conversation history, including the initial problem statement and all previous public messages. *Step 2*: the stakeholder performs role-based analysis, interpreting the available information from the perspective of their assigned role. *Step 3* (optional): the stakeholder may invoke general purpose tools (e.g. web search) and/or domain-specific tools (e.g. predictive models) to support their reasoning. *Step 4*: the stakeholder produces a public response, which may include arguments, concerns, or a new proposal. *Step 5*: this response is broadcast into the shared conversation history, making it available to all other stakeholders in subsequent turns. *Step 6* denotes the loop back to Step 1 for the next stakeholder. Once all rounds are complete, the full conversation history is passed to an LLM prompted to generate Issue/Option suggestions. **(B)** Output examples from each step. **(C)** Examples of generated Issue/Options.

To account for the human-in-the-loop approach, in a final step of the preliminary discussion the LLM-generated issues and options were presented to the workshop participants. Based on their feedback, some formulations were revised (Figure 2, step 7), thereby helping to mitigate potential LLM-driven biases or inaccuracies introduced during LLM-based issue and option generation. The adequacy of the final agreed issues and options was also evaluated in the anonymous questionnaire administered at the end of the workshop, with most participants (16 out of 18) indicating that the proposed issue-options adequately addressed the trade-offs to be considered during the negotiations (see survey link in key resources table for full details of questionnaire and answers).

### Main negotiations: LLM-based negotiation simulation for the consensus search

Each stakeholder used the final version of issue/option formulations to indicate both the importance of each issue and the option preference per issue (Figure 2, step 5). Individual scores were kept confidential from other stakeholders. Provided scores, combined with the formulation of a discussed topic, issues/options, and negotiation rules were provided as a prompt to set up an agent, virtually representing one of the stakeholders in a non-zero-sum game simulated within an LLM-based multi-agent cooperative framework [14] (Figure 2, step 6 and Figure 3 A-B; see the linked GitHub repository for the exact setup and used prompts).

Figure 4 schematically illustrates the key steps of the main negotiation simulation. In short, the negotiation simulation was implemented as a turn-based exchange between LLM-based stakeholder agents. Each agent was assigned a stakeholder-specific role, preferences, and negotiation objectives. The simulation began with an initial deal proposal, after which agents contributed sequentially according to a randomized speaking order. In each round, the selected agent analyzed the recent public conversation history, considered its own preferences, inferred the likely preferences of other stakeholders, and generated a public response that could include arguments, concessions, or a revised deal proposal. Only this public response was added to the shared conversation history (Figure 4B), while private reasoning and planning remained internal to the agent (Figure 4A, step 3). After the predefined number of rounds, the voting moderator agent generated a final proposed deal intended to balance its own objectives with the likelihood of acceptance by the other stakeholders.

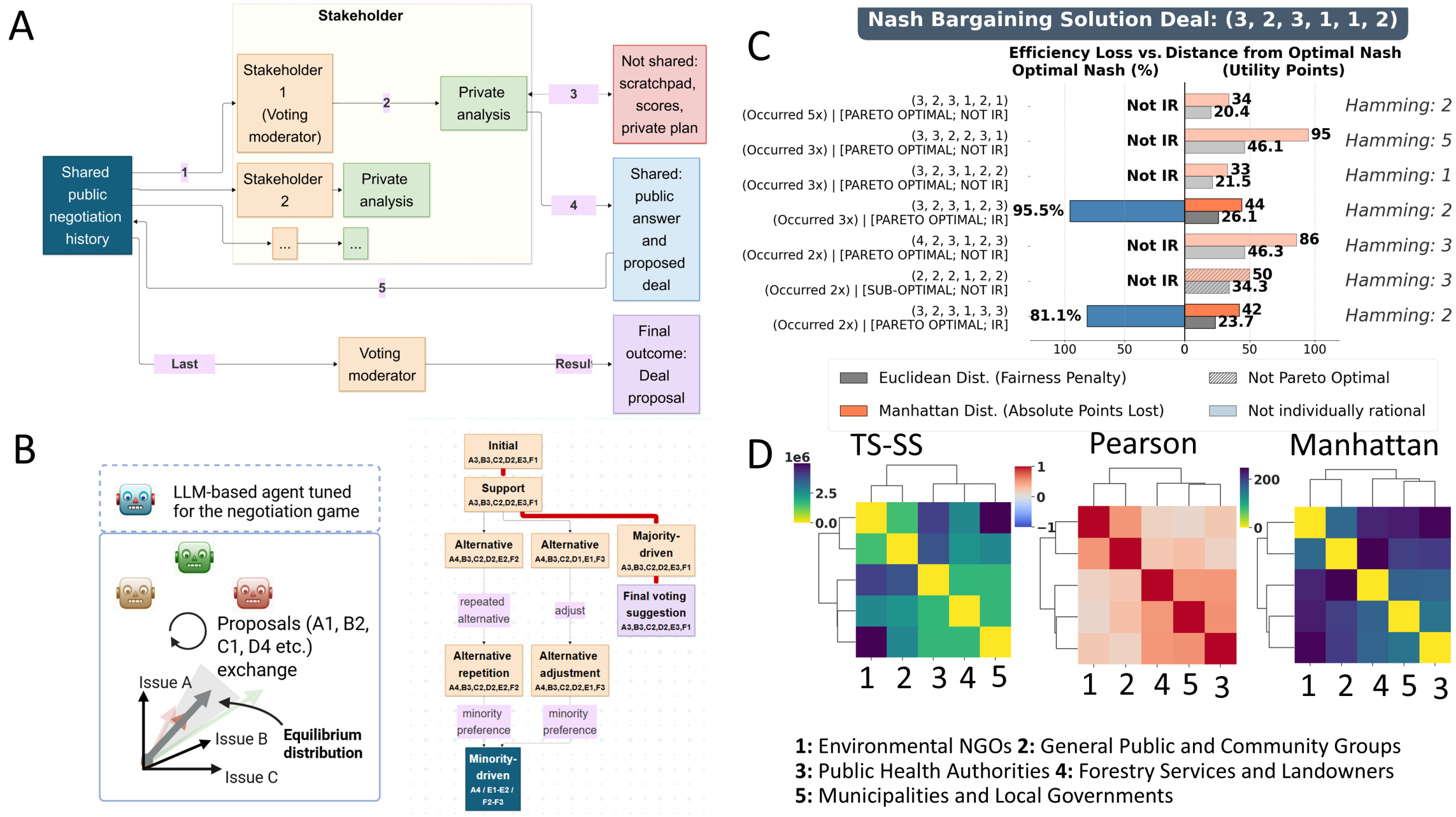

**Figure 4. Main negotiations. (A)** Schema of the main negotiation process with key steps 1-5, Last and Result. *Step 1*: A selected stakeholder receives the current shared public negotiation history, including previous public answers and proposed deals. *Step 2***:** The stakeholder is directed to the private analysis module, where it performs a role-specific analysis of the negotiation state, considering its own preferences, objectives, and the recent public exchange. *Step 3***:** Private reasoning elements-such as scratchpad content, score calculations, and planning notes-remain internal and are not shared with the other agents. *Step 4*: The stakeholder generates a public response, including its visible argumentation and a proposed deal. *Step 5***:** The public response is added back to the shared negotiation history, making it available to all stakeholders in subsequent rounds. Steps 1-5 are repeated for each stakeholder in sequence. *Last***:** After the predefined exchange rounds, the voting moderator is called. *Result:* The voting moderator produces the final outcome as a candidate deal proposal. **(B)** Example of a deal exchange among LLM-based agents, illustrating a majority-driven shift toward the final suggested deal and the emergence of competing alternatives. These details can be retrieved from the stored deal-history files. **(C)** Deviation of the most frequently proposed deals from the optimal Nash bargaining solution identified for the OPM case study, expressed as efficiency loss (%) and utility-space distance. Distances were quantified using Euclidean distance as a fairness penalty, Manhattan distance as the absolute number of utility points lost, and Hamming distance as the number of issue-level changes required to match the Nash solution. For each deal, Pareto optimality status is indicated, together with whether the deal leaves any stakeholder below their acceptance threshold, that is, whether it fails the individual rationality criterion. **(D)** Pairwise similarity between stakeholder preferences estimated using TS-SS, Pearson correlation, and Manhattan distance. Stakeholders: **1**: Environmental NGOs; **2**: General Public and Community Groups; **3**: Public Health Authorities; **4**: Forestry Services and Landowners; **5:** Municipalities and Local Governments

As with the preliminary negotiation, the main negotiation concluded with a human-in-the-loop step in which workshop participants received a report containing the distribution of the most popular deals (see S1, S3, S5, S7), and were asked to discuss them to determine whether a compromise could be reached (Figure 2, step 7). Stakeholders could also access additional information on how individual deals evolved; Figure S5) provides a detailed snapshot of one such post-analysis (see also Movie S1 for a complete post-analysis guide). As a result, in both case scenarios stakeholders were able to reach an agreement by adapting one of the most popular deals, and in the anonymous questionnaire all participants (n=18) reported that the proposed deals were useful in achieving consensus on the negotiated problem (see survey link in key resources table for full details of questionnaire and answers).

**Characterization of the LLM-suggested equilibrium outcomes**

To further increase the robustness of our results, we used several complementary metrics to characterize how optimal, from a game-theoretic perspective, the most popular deals resulting from the LLM-based negotiation simulations were (Figure 4 C). Specifically, we first assessed whether the achieved deals were Pareto optimal. The presence of multiple Pareto optimal solutions is expected in a multi-stakeholder, multi-issue negotiation; Pareto optimality was therefore interpreted as a baseline efficiency criterion rather than as evidence of global optimality. The Nash bargaining solution and Euclidean, Manhattan and Hamming distance-based metrics were used to characterize how far each deal was from the Nash-product-maximizing outcome. Euclidean distance captured overall deviation in stakeholder utility space, Manhattan distances quantified absolute point-level loss relative to the Nash benchmark, and Hamming distance expressed this deviation as the number of issue-level changes required to match the Nash solution. In addition, we evaluated whether proposed deals were individually rational with respect to the predefined minimum acceptable utility threshold (e.g. 65 utility points). Deals classified as non-individual rational (e.g. NOT IR at Figure 4) indicate that at least one stakeholder would receive less than this threshold, suggesting that this stakeholder would be left below its no-agreement baseline.

Finally, we aimed to explore possible coalitions that could form between LLM-based agents during the negotiation simulations, and used Pearson correlation, Manhattan distance, and the geometry-based Triangle Area Similarity – Sector Area Similarity (TS-SS) hybrid score to compare the pairwise similarity between stakeholder preferences (e.g., scores provided during the valuation phase, Figure 2 step 5).

In both, the OPM case (Figure 4C) and the *Campylobacter* spp. case, the disagreement-adjusted Nash bargaining solution was not achieved among the recurrent LLM-proposed deals. However, most recurrent deals were mostly Pareto optimal, indicating convergence toward locally efficient compromises rather than toward the global Nash-product optimum. The reasons for this deviation differed between the two case studies. In the *Campylobacter spp.* case, the stakeholder preference structure defined prior to the main negotiation simulation produced an outcome space in which no package satisfied the disagreement threshold for all stakeholders simultaneously. Therefore, no strict disagreement-adjusted Nash bargaining solution was defined, suggesting that at least one stakeholder would remain below its minimum acceptable utility level for every possible package.

In contrast, in the OPM case, deviation from the Nash bargaining benchmark appeared to be associated with preference alignment among a majority group of agents, which shifted recurrent proposals toward majority-preferred solutions. This potential coalition structure was further explored using stakeholder preference analysis based on TS-SS, Pearson correlation, and Manhattan distance (Figure 4D). We identified a preference-aligned cluster comprising Public Health

Authorities, Forestry Services and Landowners, and Municipalities and Local Governments. In contrast, Environmental NGOs and General Public and Community Groups formed a more distinct minority cluster. At the same time, two recurrent deals were both Pareto optimal and individually rational (Figure 4C), suggesting a possible majority-minority compromise scenario. Although these deals remained suboptimal relative to the Nash bargaining benchmark, they indicate that recurrent proposals could still reach packages acceptable for all stakeholders.

The hierarchical clustering of preference alignments (Fig. 4D) provides additional insight on how these potential coalitions may emerge. Pearson correlation and Manhattan distance placed Public Health Authorities close to the institutional-management cluster composed of Forestry Services and Landowners and Municipalities and Local Governments. In contrast, TS-SS positioned Public Health Authorities more intermediately between this institutional-management cluster and the public/environment-oriented stakeholders, namely Environmental NGOs and General Public and Community Groups. This suggests that Public Health Authorities shared some preference structure with the institutional-management stakeholders, but that this alignment was less robust when angular and magnitude-sensitive differences were jointly considered using TS-SS. Practically, this result may help identify beforehand which stakeholder within the majority-aligned group is most amenable to bridging toward the minority cluster, thereby highlighting potential directions for compromise.

**Sensitivity analysis of equilibrium characterization**

Stakeholder preferences entered the negotiation model as integer scores. Each stakeholder distributed a fixed budget of 100 points across the negotiation issues and assigned scores to the options within each issue. Because these values are judgements rather than direct measurements, the resulting game-theoretic quantities inherit uncertainty from the scoring process. We therefore examined how changes in the issue scores, together with changes in the common acceptance threshold, affected the existence and identity of the Nash bargaining solution (NBS) as well as the characterization of the recurrent deals (Figure 4C). The same perturbation model was applied to both case studies: the oak processionary moth (OPM) case and the *Campylobacter* case. However, the analyses addressed different questions. In the OPM case, an NBS existed under the provided scores, and analysis assessed the sensitivity of that solution and of the recurrent deals (Figure 4C). In the *Campylobacter* case, no package was acceptable to all stakeholders under the provided scores and the analysis therefore assessed adaptation, defined as the magnitude and configuration of possible score changes required for NBS to emerge

In short, for each stakeholder we perturbed issue scores while within-issue value ratios remained fixed, so that the perturbation altered only the relative importance of issues while preserving option orderings, ties and proportional differences within each issue. The perturbation magnitude for each issue weight was defined in the range between 1 point, reflecting the integer-valued score entries, and 25% of the original issue weight, thereby allowing proportionally larger absolute perturbations for more important issues. Perturbed issue weights were sampled jointly from the resulting constrained region using a hit-and-run Markov chain, while preserving each stakeholder's total 100-point budget. In the OPM sensitivity analysis, these samples were combined with a shared acceptance threshold perturbed symmetrically by ±10% around the baseline value of 65 points. In the *Campylobacter* adaptation analysis the common acceptance threshold was varied over a predefined grid rather than sampled randomly, allowing the effects of issue-weight adaptation and threshold reduction to be evaluated separately (see Method details for technical specifications).

For the OPM case, the NBS, individual rationality, Pareto optimality, and proximity to the NBS of the most recurrent deals were recomputed for each sample and summarized in Tables S1-S2. The deal representing the NBS under the initial scores (Figure 4C) remained the most frequently identified NBS across the perturbed samples (Table S2). Discussed most recurrent deals (Fig 4C) retained their Pareto status (e.g. optimal or not) across all samples. Among the recurrent deals that were individually rational under the original scores, the baseline Nash-efficiency losses fell within the corresponding 5th–95th percentile robustness intervals: 95.5% compared with 58.4–99.1% and 81.1% compared with 51.2–98.7% (Table S1, R3 and R6 ). This indicates that the baseline estimates were consistent with, rather than outliers relative to, the distributions obtained under score and threshold perturbation.

At the same time, the existence alternative NBS packages (frequencies from 0.1 to 18.6%, the original NBS remained in 25% of samples) (Table S2) as well as the shift of two previously not individually rational deals to the group of individually rational compromises (Table S1; R1 and R2) further support the exploration of fair compromises during the negotiation process when the original NBS is not selected.

For the *Campylobacter* case, the adaptation analysis identified two closely related deals, with a Hamming distance of one, that emerged as NBS solutions under different combinations of preference perturbations and acceptance thresholds (Fig S11).The analysis thereby identified the specific preference and threshold changes under which agreement became possible and revealed a pair of closely related compromise packages, although only one package was feasible at the original acceptance threshold of 65. In practical applications, such an adaptation analysis could be performed after the scoring step (Figure 2, step 5) and before the main negotiations are launched (Figure 2, step 6), helping to identify promising compromise packages and guide the negotiation process towards a fair and feasible agreement.

The sensitivity analysis verified that the deal characterization reported in Figure 4 is stable with respect to both the scoring procedure (Figure 2, step 5) and the acceptance-threshold setting originally taken from Abdelnabi et al (2023). This check is essential, because elicited scores are point estimates of latent preferences that inevitably carry the subjectivity of the stakeholders who provided them as well as the granularity of the scoring procedure, while the acceptance threshold is an external constant adopted from the original protocol rather than a quantity elicited from the participants or derived from the cases themselves. It should be noted, however, that the perturbation was applied downstream of the negotiations: the deals themselves were generated by LLM-based simulations that used the original, unperturbed scores. We therefore cannot exclude the possibility that a different set of deals would emerge if the simulations were re-run under perturbed preference landscapes. Because the present approach was designed as a human-in-the-loop application rather than a standalone system, we did not pursue this additional degree of variation. We acknowledge, however, that this restricts both the transferability of our conclusions to fully automated deployments and our ability to isolate the contribution of the LLM-based simulations themselves, since the deals and the preference landscape against which they are assessed derive from a single, unperturbed set of elicited scores.

## Discussion

Tackling risks can only be done in an interdisciplinary and multidimensional manner, where stakeholders from various disciplines use their expertise to negotiate risks and balance trade-offs. However, such multidisciplinary negotiations will inevitably face difficulties, including the lack of a common currency to assess risks and /or benefits and the challenge of balancing all stakeholders' interests under time constraints [9]. In addition, during the risk assessment process, data gaps often become apparent, and the generation and consideration of new data may be needed to minimize uncertainty. In the present work, we combined

RAG-based information extraction [11], LLM-based agentic discussion simulation accompanied by domain-specific tool calls [30], and a “human-in-the- loop” (HIL) approach to augment negotiation-centered risk analysis. Specifically, we leveraged LLMs (i) to help stakeholders formulate the specific issues (e.g. risks) to discuss, as well as the possible risk management options to pursue, and (ii) to propose agreements aimed at achieving a balance that maximizes overall benefits while minimizing risks. In both case scenarios, stakeholders completed the risk negotiation within the allotted time, demonstrating the feasibility of the workflow under realistic workshop conditions. However, agreement was reached only after participants discussed and modified the LLM-proposed deals, and no unassisted control condition was included. The specific contribution of our LLM-based pipeline to the negotiation process and its outcome therefore cannot be isolated (see limitations).

Information exchange among stakeholders from various disciplines is considered essential for effective risk analysis [31,32]. In this regard, the simulated preliminary discussion on a given topic prior to the issue and option specification is of particular importance. Specifically, we used LLMs to derive issue and option formulations not solely based on individual proposals from each stakeholder, but rather on the full text of the simulated preliminary discussion. Since this discussion incorporates stakeholder positions grounded in the broad scope of the provided scientific literature, along with their motivations, arguments and domain-specific information retrieved through tool calls, as well as their expressed willingness to adapt their opinions, we hypothesize that the resulting issue and option formulations may be more semantically enriched and capture more nuances than those derived from direct stakeholder proposals alone; this was not tested here and warrants comparative evaluation. Additionally, our pipeline aims to identify a compromise among stakeholders’ perspectives but does not independently verify the data underpinning their opinions, nor does it draw on stakeholder expertise beyond what is encoded in the documents used for RAG-based inference or in the directly completed agent descriptions. In this context, the preliminary discussion-in which LLMs simulate an exchange of stakeholder opinions and integrate supplementary information via targeted web searches- may serve as an important safeguard, potentially allowing stakeholder positions to be partially validated by cross-referencing stated facts with publicly accessible information.

Our risk negotiation simulation is built on an LLM-based non-zero-sum negotiation game setup [14]. We devised a process for structuring issue/option formulation and subsequent preference scoring and outlined a procedure for HIL review of the obtained results. We propose that such formalization can help to ensure that the interests of all stakeholders are equally accounted for. The issue score serves as an indicator of the relative importance of each issue, while the distribution of scores across the various options linked to that issue signals potential areas of compromises or their absence. Stakeholders are also allowed to modify the option formulations at the final negotiation stage and may choose to disclose parts of their confidential information such as maximum scores per issue or their complete risk valuation-to ease the compromise search. Finally, our approach is not bound to a specific LLM such as GPT-4o [22] or a specific framework, such as Langchain [33], and the presented solution is one of many possible implementations of the proposed framework. We consider the main added value of this work in the procedural linkage between the negotiation framework – the six-step approach described by Ehling-Schulz *et al.* [9] - and computational science methods applied here.

Together, these findings suggest that LLM-based negotiation simulations can identify locally efficient compromise solutions but may not reliably converge toward globally balanced bargaining outcomes. The frequent occurrence of Pareto-optimal deals indicates that the agents were often able to avoid clearly inefficient proposals. However, deviations from the disagreement-adjusted Nash bargaining benchmark highlight two important limitations. First, when stakeholder preferences are structurally incompatible, as observed in the *Campylobacter spp.* case, no package may satisfy all minimum acceptability thresholds simultaneously. Second, even when individually rational compromises exist, as in the OPM case, the negotiation process may still be affected by majority-aligned preference clusters, potentially leaving minority stakeholders with less favorable but still acceptable outcomes. Thus, the proposed framework is useful not only for identifying candidate compromises, but also for diagnosing why potential outcomes may not be "fair" in the mathematical sense. In practical applications, this type of analysis could help reveal whether disagreement arises from an inherently infeasible preference structure or from coalition-like preference alignment that shifts negotiations toward majority-preferred solutions.

Being a moderator confers the advantage of being able to start and conclude the conversation with a preferred option. Therefore, the moderating side has an additional indirect weight in its position, which adds a new dynamic to the negotiation process and may help to prevent it from settling for the local minima defined by the opposing side(s). Getting stuck in local minima is a common issue in real-world optimization problems and is an area of active research [34]. Therefore, here we additionally pinpoint an important degree of freedom that should not be neglected during the negotiation flows. Further, we highly recommend appointing an experienced moderator with strong skills in facilitation and discussion management, as well as impartial expertise in the field, to effectively guide the discussions during real negotiation exercises. Additionally, stakeholders could use the archived simulated conversations to see how the discussion evolved, what strategies were effective and which discussion points remained unsolved, thereby potentially pointing to areas of compromise and divergence. This information is of particular importance, if the proposed pipeline is used as preparation for the actual negotiations. When discussing the results of the simulated negotiation, stakeholders stressed the issues and options which they were adamant about and, in turn, signaled a willingness to compromise on other issues. In both exercises, stakeholders reached consensus through this human discussion and modification of the proposed deals, reflecting the integrating of a HIL approach into our process, to ensure that minority opinions are accounted for.

The negotiation dynamics observed in our case scenarios can be anchored in established theoretical frameworks. The moderator's structural advantage in shaping opening and closing proposals is consistent with anchoring effects and agenda-setting power documented in behavioral negotiation research [35]. The coalition dynamics — where two stakeholders with aligned interests exerted collective pressure on the third — align with multi-party coalition theory [36], and underscore why the HIL approach is essential to safeguard minority positions [37]. Finally, stakeholders' voluntary partial disclosure of confidential scores mirrors strategic information revelation in principled negotiation, facilitating identification of the zone of possible agreement [37]. That dynamics consistent with these frameworks emerged in our two exercises suggests that the framework may capture substantive negotiation mechanisms beyond procedural simulation; this qualitative observation requires systematic confirmation.

## Limitations

The present study did not include matched non-LLM negotiation condition. Consequently, the observed completion of both negotiation exercises within the allocated workshop time and participants' positive evaluations should be interpreted as evidence of feasibility of the proposed approach and not directly as evidence that LLM assistance shortens negotiation time, increases the probability of agreement, or improves the quality of the resulting decisions. Moreover, because of our HIL-set-up, the final agreements emerged from human discussion and modification of LLM-proposed deals, the effects of LLM assistance are confounded with those of moderated human deliberation. In addition, participant evaluations were self-reported by a small sample (n=18) after a single workshop. Establishing such comparative effects will require controlled studies in which equivalent negotiation problems are conducted with and without LLM assistance under matched participant background, information availability, and time constraints, using predefined outcome measures such as time to agreement, agreement rate, and expert-rated decision quality, and the extent to which final agreements deviate from LLM-proposed deals.

As our framework relies on LLMs, it is limited by the fundamental issues linked with their usage such as hallucinations, quantitative inaccuracies, lack of reasoning abilities, and biases. LLMs can generate plausible-sounding but incorrect or nonsensical information, often due to the quality of their training data and the probabilistic nature of their text generation processes [38]. LLMs can struggle with mathematical tasks [39], and be prone to biased conclusions derived from the unfiltered training datasets containing stereotypes, misrepresentations, social/ethical biases etc. [40] Even though there are recent developments mitigating such limitations [41,42], we argue that a human-in-the-loop approach is still essential to address such risks associated with LLMs in accordance with [43] . Yet, in a rapidly evolving AI domain such as LLM-based modelling, one may foresee dramatic shifts towards independently interacting and negotiating agents without human oversight. The described higher performance of GPT-4 alone compared to physicians alone, and/or in combination with GPT-4 [44], was one of the first benchmarking results towards this shift. Importantly, our negotiation simulation pipeline can, also, be run in an autonomous manner [15] to enable the exploration of the variety of possible negotiation scenarios before the actual negotiations. From an ethical perspective, when using LLMs in risk negotiation, we need to decide how far risk negotiation should be influenced by LLMs and what the human corrective is to assure diversity, equity and the protection of the weak [45].

Within the current work, we focused on a cooperative negotiation scenario where all stakeholders seek a compromise, whereas the reach dynamic of multi-agent communication allows researchers to simulate other plausible situations such as sabotaging or greedy games scenarios [14]. Another development direction is to enrich LLM-based agents with personality traits (e.g. extraversion, neuroticism, etc. [46,47]) that greatly influence real-case negotiations [48]. Finally, the integration of the next-generation LLMs with advanced complex reasoning features such as GPT-5 family [49] may further improve the quality of the simulated negotiations, which should be assessed empirically. It was shown that LLMs can help humans find common ground in democratic deliberation [50]. The negotiating agents in our framework can also be augmented with an additional layer that learns the interests of the other agents from the propositions. Different methodologies can be used, as described by [51] for bilateral negotiations.

Our implementation relies on role-based prompting and a shared dialogue history and therefore does not constitute a fully autonomous multi-agent system. Benchmarking against alternative approaches, and more autonomous agent architectures remain directions for future work.

**Conclusion**

Our AI-assisted negotiation-centered risk analysis approach proved feasible in a workshop with independent domain experts across two real-world One Health scenarios, in which participants reached consensus within the allotted time after discussing and modifying the LLM-proposed deals. As the contribution of LLM assistance to this outcome cannot be isolated, whether it mitigates the time and information constraints stakeholders experience during roundtable discussions remains to be established in controlled comparative studies. We believe that our approach offers new opportunities to integrate trade-offs in risk analysis and could support more balanced, effective strategies[52]. It may also be used to simulate potential negotiation scenarios, offering stakeholders an insight into how the negotiation process might unfold. At the same time, having human supervision at all principal preparatory stages enables monitoring for potential data inherent biases and alignment with ethical principles. If confirmed on comparative evaluations, such approaches could contribute to more efficient and effective policymaking and management of risks and help tackle several of the pressing challenges facing our society.

## Resource availability

### Lead contact

Requests for further information should be directed to and will be fulfilled by the lead contact, Sophia Johler (Sophia.Johler@lmu.de).

### Materials availability

This study did not generate any unique reagents.

### Data and code availability

This paper does not report datasets of a standardized data type (e.g., sequencing, proteomics, and crystallography). Custom-formatted data generated for this study along with the codes are deposited in the project related repository https://github.com/KIDA-BfR/Risks-assessments-LLMs

## STAR★Methods

### Key resources table

| Reagent or Resource | Source | Identifier |
| --- | --- | --- |
| **Deposited Data** | | |
| Copy of the codes used for the initial development | Zenodo | https://zenodo.org/records/17095804 |
| Anonymous survey from October seminar | Github | https://github.com/KIDA-BfR/Risks-assessments-LLMs/tree/main/Revision_1/October_seminar_logs/Seminary_Survey |
| **Software and algorithms** | | |
| Application used during the seminar | Github | https://github.com/MSG4BFR/negotiation-workshop-app |
| Codes for the simulation of the preliminary discussion and main negotiations | Github | https://github.com/KIDA-BfR/Risks-assessments-LLMs |

**Methods details**

**Proof-of concept development and testing**

To develop an initial proof-of-concept practical solution for the negotiation-centered risk analysis framework, two case scenarios were used, each highlighting One Health considerations. The first case scenario relates to the use of *Bacillus (B.) thuringiensis* as a biopesticide in agriculture, involving negotiating tradeoffs between using *B. thuringiensis* as a biopesticide to reduce transmission of insect borne pathogens and a risk of *B. thuringiensis* deposited on crops potentially causing foodborne illness ("case scenario 1: *Bti*"). The second case scenario addresses tradeoffs between hunting of wild boars as population control option, to lower crop damage, biodiversity loss, and disease spread, and alternative population control methods to minimize animal suffering ("case scenario 2: hunting").

For the initial development, project team members assumed roles representing one of three stakeholder groups in each scenario. The team members who participated in the above-described scenario were familiar with the scientific background of the cases and could therefore confirm the scientific relevance of the obtained results, which was important during the proof-of-concept development phase. To simulate round-table discussions, two virtual exercises were conducted per scenario. All participants received clear instructions beforehand and were asked to use their expertise to provide specific information needed to generate agent descriptions. Each virtual exercise session was set for two hours and moderated objectively by another project team member to guide discussions. These exercises were held between May and August 2024.

### Evaluation of the LLM-based risk-negotiation approach through real-stakeholder workshop set-up

For evaluating the practical utility of the developed approach, we organized a real-case scenario workshop involving experts in food microbiology, environmental contaminants, and risk assessment from academia, government, and industry. Prior to the workshop, participants were asked to share materials - such as scientific publications, field reports, or other published documents - that were considered relevant for defining the agent descriptions. Additional material was retrieved through classical literate evidence search in Pubmed and through AI-based search engines, e.g. ChatGPT[53] . All relevant materials were subsequently used to generate agent descriptions using RAG-based methods [11].

The workshop was organized as a joint on-site event of the RAKIP-Initiative [13] and MicRISK2030 project consortium [54], and was held at the German Federal Institute for Risk Assessment (BfR) in Berlin in October 2025 [55]. During the workshop, participants engaged in two real-world case study negotiation exercises. The first case study addressed the question: “How best to balance trade-offs for setting a Food Safety Criterion for *Campylobacter* spp. in broiler meat under EU legislation 2073/2005?” with simulated stakeholders being: (i) Food Safety Authorities, (ii) Food Industry, (iii) Environmental NGOs, and (iv) Consumers (case scenario 3:

*Campylobacter spp.*). The second case study posed the question:” Should the management of the oak processionary moth (OPM) in Central Europe prioritize public health measures over ecological conservation, considering the potential health risks, environmental impact, and economic costs involved?” The simulated stakeholders for this scenario were: (i) Municipalities and Local Governments, (ii) Public Health Authorities, (iii) Forestry Services and Landowners, (iv) Environmental NGOs, and (v) General Public and Community Groups (“case scenario 4: OPM”). A human moderator guided participants through all phases of application and the discussion according to predefined guidelines (see Supplementary Text S11).

To ensure effective use of the developed pipeline by the large number of participants, we designed a web application (see Key Resource Table) that allowed participants to provide feedback on the results obtained during the preliminary negotiation phase, submit preference scores, and view the final negotiation results during the main negotiation stage. The LLM-proposed candidate deals were subsequently discussed by participants in a moderated session, during which participants could modify individual options within a deal, or align on a combination of the most frequent LLM-proposed deals.

Additionally, participants of this workshop were asked to provide anonymous feedback on the utility of each step of the approach through a survey, which was set-up using SurveyMonkey® and administrated at the end of the workshop (see survey link in key resources table for full details of questionnaire and answers).

## Analysis of main negotiation results

### *Package utility and outcome spaces*

Let $S$ denote the set of stakeholders and $I$ denote the set of negotiation issues:

$$
\begin{aligned}
S &= \{1,2,\dots,n\},\\
I &= \{1,2,\dots,m\}.
\end{aligned}
$$

Each issue $i \in I$ has a finite set of possible options $O_i$. A complete negotiation package is defined as one selected option for every issue:

$$p = (o_1, o_2, \dots, o_m), \qquad o_i \in O_i.$$

The full outcome space is the Cartesian product of all option sets:

$$\mathcal{P} = O_1 \times O_2 \times \cdots \times O_m = \prod_{i=1}^{m} O_i.$$

For each stakeholder $s \in S$, the option score for choosing option $o_i$ under issue $i$ is denoted by $v_s(i, o_i)$. Stakeholder-specific package utility was computed as the additive sum of option scores across all issues:

$$U_s(p) = \sum_{i=1}^{m} v_s\,(i, o_i).$$

Thus, each package $p$ is represented by a stakeholder utility vector:

$$\mathbf{U}(p) = \big(U_1(p), U_2(p), \dots, U_n(p)\big).$$

***Pareto optimality***

Pareto optimality was used as a baseline efficiency criterion. A package $q \in \mathcal{P}$ Pareto-dominates another package $p \in \mathcal{P}$ if all stakeholders receive higher or equal utility under $q$, and at least one stakeholder receives higher utility:

$$q \succ_P p[\forall s \in S,\ U_s(q) \geq U_s(p)] \wedge [\exists s \in S,\ U_s(q) > U_s(p)].$$

A package is Pareto optimal if no other package in the full outcome space dominates it. The Pareto set is therefore:

$$\mathcal{P}^* = \{p \in \mathcal{P} \mid \nexists q \in \mathcal{P}: q \succ_P p\}.$$

If a negotiated package is not Pareto optimal, then at least one alternative package exists that would improve the utility of one or more stakeholders without reducing the utility of any other stakeholder. In this sense, non-Pareto packages indicate unrealized attainable value within the structured outcome space.

***Disagreement-adjusted Nash bargaining benchmark***

A Nash-product benchmark was computed for all packages in the outcome space. In contrast to a zero-disagreement formulation, the disagreement point was set to 65 utility points for every stakeholder. This means that if no agreement is reached, each stakeholder is assumed to receive a fallback utility of 65. Therefore, only the utility gains above this disagreement baseline contribute to the Nash product.

$$d_s = 65, \qquad \forall s \in S$$

For a package $p$, the disagreement-adjusted utility gain for stakeholder $s$ is:

$$G_s(p) = U_s(p) - d_s$$

A package was considered individually rational if all stakeholders received at least their disagreement utility:

$$p \in \mathcal{P}^{IR} \Leftrightarrow \forall s \in S,\ U_s(p) \geq d_s$$

where $\mathcal{P}^{IR}$ denotes the set of individually rational packages. The disagreement-adjusted Nash product was then defined as:

$$NP_{d_s}(p) = \prod_{s=1}^{n} [U_s(p) - d_s], \qquad p \in \mathcal{P}^{IR}$$

The Nash bargaining solution was identified as the individually rational package that maximized this disagreement-adjusted Nash product:

$$p_{d_s=65}^{\text{NBS}} = \arg\max_{p \in \mathcal{P}^{IR}} NP_{d_s}(p) = \arg\max_{p \in \mathcal{P}^{IR}} \prod_{s=1}^{n} [U_s(p) - d_s]$$

If no package satisfies the individual-rationality condition, then no disagreement-adjusted Nash bargaining solution exists under the assumed disagreement point of 65:

$$\mathcal{P}^{IR} = \varnothing \Rightarrow p^{\mathrm{NBS}}_{d_s=65} \text{ does not exist}$$

Importantly, $p^{\mathrm{NBS}}_{65}$ should be interpreted as a Nash bargaining-type benchmark relative to the disagreement baseline of 65, rather than as a strategic Nash equilibrium based on unilateral deviations. Compared with the zero-disagreement formulation, this version asks which feasible package maximizes the product of stakeholders' gains over their fallback outcome, not simply the product of their total utilities.

***Distance of negotiated outcomes from the Nash bargaining solution***

Let $p_r$ denote the package selected in negotiation round $r$, and let $p^{\mathrm{NBS}}$ denote the Nash bargaining solution. The distance between the observed deal and the Nash benchmark was calculated using several complementary metrics.

***Manhattan distance in utility space***

The Manhattan distance was used to measure the aggregate absolute deviation in stakeholder utilities from the Nash bargaining solution:

$$D_1(p_r, p^{\mathrm{NBS}}) = \sum_{s=1}^{n} \left| U_s(p^{\mathrm{NBS}}) - U_s(p_r) \right|.$$

This quantity captures how many utility points, summed over stakeholders, separate the selected deal from the Nash benchmark.

***Euclidean distance in utility space***

The Euclidean distance was calculated as:

$$D_2(p_r, p^{\mathrm{NBS}}) = \sqrt{\sum_{s=1}^{n} \left( U_s(p^{\mathrm{NBS}}) - U_s(p_r) \right)^2}.$$

Because deviations are squared before summation, this metric is more sensitive to large stakeholder-specific gaps. Therefore, it can reveal whether deviation from the Nash benchmark is concentrated among one or a few stakeholders rather than evenly distributed across all stakeholders.

### *Nash percentage loss*

The Nash percentage loss measured the proportional decrease in the Nash product relative to the Nash bargaining solution:

$$L_{\mathrm{Nash}}(p_r) = \frac{NP(p^{\mathrm{NBS}}) - NP(p_r)}{NP(p^{\mathrm{NBS}})} \times 100.$$

This metric summarizes how much the selected deal falls short of the Nash-product-maximizing package. Because the Nash product is multiplicative, this loss is sensitive both to aggregate inefficiency and to strongly imbalanced utility distributions.

### *Hamming distance in issue space*

Utility-space distances were complemented by an issue-level comparison. If $o_{r,i}$ is the option selected for issue $i$ in round $r$, and $o_i^{\mathrm{NBS}}$ is the corresponding option in the Nash bargaining solution, the Hamming distance was defined as:

$$D_H(p_r, p^{\mathrm{NBS}}) = \sum_{i=1}^{m} \mathbf{1}\left(o_{r,i} \neq o_i^{\mathrm{NBS}}\right).$$

This metric counts how many issue-level choices differ from the Nash bargaining solution. It therefore translates utility loss into practical negotiation friction by estimating how many agenda items would need to be revised for the selected deal to match the Nash benchmark

### *Coalition exploration using stakeholder preference vectors*

Potential coalition structures were explored by comparing stakeholders' full preference profiles across all issue-option scores. Each stakeholder $s$ was represented as a vector of option scores. Pairwise stakeholder similarity or distance was then computed using Pearson correlation, Manhattan distance, and TS-SS distance.

### *Pearson correlation and Manhattan distance*

Pearson correlation was used to evaluate whether two stakeholders' preference profiles varied in a similar direction across options, independently of absolute score magnitude. High positive correlation indicates that two stakeholders tend to prefer and disprefer similar options, even if their absolute score levels differ.

The Manhattan distance between two stakeholder preference vectors captures the total absolute difference in raw option scores and is therefore sensitive to the magnitude of preference disagreement.

### *TS-SS distance*

The TS-SS metric was used as an additional geometric comparison that combines angular and magnitude-sensitive components. For two stakeholder preference vectors $\mathbf{x}_s$ and $\mathbf{x}_t$, cosine similarity was first computed as:

$$\cos(\theta_{s,t}) = \frac{\mathbf{x}_s \cdot \mathbf{x}_t}{\| \mathbf{x}_s \| \| \mathbf{x}_t \|}.$$

In the implementation, the angular component was shifted by 10 degrees:

$$\theta'_{s,t} = \arccos\left(\cos(\theta_{s,t})\right) + 10^\circ.$$

The Euclidean distance and magnitude difference were calculated as:

$$\begin{aligned} E_{s,t} &= \sqrt{\sum_{k=1}^{K} \left(x_{s,k} - x_{t,k}\right)^2}, \\ M_{s,t} &= \left| \| \mathbf{x}_s \| - \| \mathbf{x}_t \| \right|. \end{aligned}$$

The triangle component was defined as:

$$TS_{s,t} = \frac{\parallel \mathbf{x}_s \parallel \parallel \mathbf{x}_t \parallel \sin(\theta'_{s,t})}{2}.$$

The sector component was defined as:

$$SS_{s,t} = \pi(E_{s,t} + M_{s,t})^2 \frac{\theta'_{s,t}}{360^\circ}.$$

The final TS-SS distance was:

$$D_{\text{TS-SS}}(s,t) = TS_{s,t} \times SS_{s,t}.$$

This measure jointly accounts for differences in vector direction and vector magnitude and therefore complements Pearson correlation and Manhattan distance.

### *Hierarchical clustering and coalition interpretation*

For each pairwise distance matrix, hierarchical clustering was performed using average linkage. For clusters $A$ and $B$, average linkage defines inter-cluster distance as:

$$D(A,B) = \frac{1}{|A||B|} \sum_{s \in A} \sum_{t \in B} D\,(s,t).$$

Stakeholders placed close together in the resulting dendrogram were interpreted as having similar preference profiles and therefore as potential candidates for the coalition formation. However, these clusters should be interpreted as preference-based coalition hypotheses rather than direct evidence of coordinated strategic behavior. Actual coalition formation would require additional evidence from negotiation discourse, voting behavior, or explicit strategic alignment among stakeholders.

### *Sensitivity and adaptability analysis*

Sensitivity and adaptability analysis was designed to systematically perturb issue scores to examine the stability of NBS and parameters of the most recurrent deals for the OPM case, as well as to identify the possible NBS for the *Campylobacter spp.* case. 25% issue score perturbation range was chosen, since for the *Campylobacter* case the NBS was found under the allowed perturbation of the issue scores at 22% and we aimed to keep the perturbation comparable in both case scenarios. 10% threshold variation was selected to have a comparable absolute unit perturbation between the max issue score variation in the OPM case (e.g. 6.75) and the variation of the acceptance threshold (e.g. 6.5 points). At the next sections we provide additional details of such procedures, with all corresponding codes available in GitHub.

***Perturbation strategy***

Let $v_{sio}$ be the score assigned by stakeholder $s$ to option $o$ of issue $i$. S denote stakeholders, J issues and $\mathcal{O}_i$ the options available for issue I as it was *"Package utility and outcome spaces"* section. The weight of issue $i$ and the total scoring budget (T) of stakeholder $s$ are

$$w_{si} = \max_{o \in \mathcal{O}_i} v_{sio}, \qquad T_s = \sum_{i \in \mathcal{I}} w_{si} = 100.$$

To separate issue importance from within-issue preference, the relative value of an option was defined as

$$\rho_{sio} = \begin{cases} v_{sio}/w_{si}, & w_{si} > 0, \\ 1, & w_{si} = 0. \end{cases}$$

Thus, $\rho_{sio} \in [0,1]$ records the proportion of the issue weight delivered by option $o$. When an issue was originally assigned zero weight, all its options were tied at zero. Setting $\rho_{sio} = 1$ for every option preserves that indifference if the issue is subsequently given a small positive weight: the utility level changes equally for all packages, but their ranking is unaffected by that issue.

For a perturbed issue-weight vector $w'_s = (w'_{s1}, \dots, w'_{sm})$, the additive utility of package $d$ is

$$U_s(d; w'_s) = \sum_{i \in \mathcal{I}} \rho_{sid_i}\ w'_{si}.$$

Thus, the perturbation changes only the relative importance of the issues. The ordering, proportional spacing, ties, and structural zeros among options within each issue remain exactly as elicited.

For stakeholder $s$ and issue $i$, the perturbation tolerance was

$$c_{si}(\gamma) = \max\{\kappa, \gamma w_{si}\},$$

where $\kappa = 1$ point reflecting asked integer inputs for scores and $\gamma = 0.25$ reflecting the allowed variation

The lower and upper bounds were

$$\ell_{si} = \max\{0, w_{si} - c_{si}(\gamma)\}, \qquad u_{si} = w_{si} + c_{si}(\gamma).$$

For an issue with $w_{si} = 0$, the admissible interval was set directly to $[0,1]$.

Therefore, the complete set for stakeholder S was

$$\mathcal{W}_s(\gamma) = \left\{ w'_s\colon\ \ell_{si} \le w'_{si} \le u_{si}\ \forall i,\quad \sum_{i \in \mathcal{I}} w'_{si} = T_s \right\}.$$

The equality constraint preserves the 100-point scoring. Consequently, every increase in the importance of one issue must be offset by decreases elsewhere and uniform draws from each $\mathcal{W}_s(\gamma)$ were therefore generated using a hit-and-run Markov chain.

Additionally, we have independently varied the acceptance threshold of 65 which was originally chosen to reproduce the prior work from Abdelnaby et al. (2023) for cooperative scenarios.

Let $r$ denote the common minimum acceptable utility. A package $d$ is therefore acceptable for the stakeholder if

$$U_s(d; w'_s) \geq r \qquad \forall s \in \mathcal{S}.$$

In the OPM sensitivity analysis, threshold uncertainty was represented by one shared draw for all stakeholders,

$$r' = 65(1 + z), \qquad z \sim \text{Uniform}(-0.10, 0.10)$$

Using a shared value preserves the experimental rule that all stakeholders face the same acceptance threshold.

### *Adaptation analysis for the Campylobacter spp. case*

In the *Campylobacter spp.* case, the issue-score perturbation followed the perturbation strategy described above and issue weights varied within $c_{si}(\gamma) = \max\{1, \gamma w_{si}\}$, with the 100-point scoring and within-issue option preferences preserved. However, it was used for adaptation rather than sensitivity analysis because no NBS existed under the original scores.

As in the *Campylobacter spp.* adaptation analysis we sought conditions where NBS emerges, $r$ was varied over a controlled grid so that the effects of changing the threshold could be distinguished from the effects of reallocating issue weights.

## Prompt engineering details

Below we provide the overall structure of the prompt engineering strategy used for the preliminary and main negotiation simulations, while the GitHub repository from key resource table contains the exact prompts that were used during all scenarios

### *Preliminary negotiations*

The negotiation pipeline used a role-conditioned prompt engineering strategy to instantiate each stakeholder as an LLM-based agent. First, stakeholder-specific descriptions converted into system prompts containing the agent's name, perspective, and argumentative goal. Each agent was instructed to persuade other participants from its assigned viewpoint, use external tools to verify or challenge claims, cite sources when tool calls were used and avoid fabricating references. Additionally, for *Campylobacter spp.* case we provided explicit reference numbers of the web-hosted microbiological computational models to be used[28,29]. During the simulated discussion, agents did not operate with independent strategic learning rules, but instead, their behavior was guided by their role prompt, tool access, and the shared dialogue history accumulated across turns. A moderator prompt initialized the discussion by specifying the overall negotiation topic and asking participants to formulate mutually acceptable issues and options. The resulting exchange was then summarized by a separate prompt into structured issue–option formulations suitable for the later preference elicitation and negotiation-game stages.

### Main negotiations

To guide the negotiation agents, we used the following prompt engineering approach from Abdelnabi et al. [56] . In short, the prompts enforce a "secret" Chain-of-Thought process by requiring agents to articulate intermediate calculations, assess other parties' preferences, and formulate strategies within a text inside XML-style delimiters <SCRATCHPAD> before generating a public response. To maintain continuity across negotiation rounds, the framework implements forward planning and memory injections: agents draft private notes (<PLAN>) detailing future options, which are subsequently injected into their prompts in the following rounds alongside instructions to adapt these plans based on new observations. Furthermore, to manage token limits and maintain contextual focus, a sliding context window is utilized, presenting the agents with a personalized, chronological history of up to a predefined number of recent interactions. To prevent breakdowns, explicit behavioral guidelines are integrated into the prompt, strictly instructing the agents to prioritize balanced agreements, empathize with counterparts, and seek win-win outcomes as long as their minimum score (e.g. 65) is met. Finally, the system relies on strict delimiters (e.g., <ANSWER>, <DEAL>, <PLAN>) to ensure the generated textual outputs are cleanly parsed by the execution script and accurately routed to either the public negotiation history or the agent's private state.

### Funding statement

This work was funded by the SNSF (Swiss National Science Foundation, Sinergia grant number CVRSII5_209253).

## Declaration of competing interest

None declared.

## Supplementary Material

***The PDF file includes:***

Supplementary Text S1-S11

Supplementary Figures S1-S11

Supplementary Tables S1-S2

***Other Supplementary Material for this manuscript includes the following:***

Movie S1

## Declaration of generative AI and AI-assisted technologies in the manuscript preparation process

During the preparation of this work, which is about Large Language Models applicability for risk negotiation, the author(s) used several generative AI and AI-assisted technologies in order to prepare the manuscript. After using this tool/service, the author(s) reviewed and edited the content as needed and take(s) full responsibility for the content of the published article.

# Supplementary Material

## Towards using Large Language Models for Risk Negotiation and Consensus-building in One Health contexts

Alexandra Fetsch, Iurii Savvateev, Racem Ben Romdhane, Martin Wiedmann, Artemiy Dimov, Maciej Durkalec, Josef Teichmann, Jakob Zinsstag, Konstantinos Koutsoumanis, Andreja Rajkovic, Jason Mann, Mauro Tonolla, Monika Ehling-Schulz, Matthias Filter, Sophia Johler

Corresponding author: Prof. Sophia Johler
Email: sophia.johler@lmu.de

**This PDF file includes:**

Supplementary Text S1: Description and background information of case scenarios 1-4.
Supplementary Text S2: Description of the steps of the AI-assisted negotiation-centered risk analysis process.
Supplementary Text S3: Template Scoring Case scenario 1 (*B. thuringiensis*).
Supplementary Text S4: Template Scoring Case scenario 2 (wild boar hunting).
Supplementary Text S5: Final Issues/Options Case scenario 3 (*Campylobacter spp*).
Supplementary Text S6: Final Issues/Options Case scenario 4 (OPM).
Supplementary Text S7: Case scenario 1: Full details of negotiation outcomes.
Supplementary Text S8: Case scenario 2: Full details of negotiation outcomes
Supplementary Text S9: Case scenario 3: Full details of negotiation outcomes.
Supplementary Text S10: Case scenario 4: Full details of negotiation outcomes.
Supplementary Text S11: Moderator instructions during evaluation workshop.

Supplementary Fig. S1, S3, S6, S8: Combined results – distribution of unique issue-option combinations.
Supplementary Fig. S2, S4, S7, S9: Moderator influence on the issue-option combinations.
Supplementary Fig. S5: Overview of the additional data to track the deals history
Supplementary Fig. S10: Coalition analysis for *Campylobacter spp.* case
Supplementary Fig. S11: Emergence of NBS in Campylobacter spp. case during adaptation analysis

Supplementary Table 1: Sensitivity analysis for OPM case, matched with values from Fig. 4C
Supplementary Table 2: Distribution of NBS deals during sensitivity analysis

**Other supplementary materials for this manuscript include the following:**

Movie S1: Visualization of the negotiation simulation

This video is related to step 4 (risk negotiation) of the AI-assisted negotiation-centered One Health risk analysis procedure.

**Supplementary Material**

**Supplementary Text S1**

**Description and background information of case scenarios used for development and testing**

**Case scenario 1: *Bacillus thuringiensis***

*B. thuringiensis* is a naturally occurring bacterium present in soil, and the environment, that among others produce a set of proteins toxic for a number of insects and, thus, can be used as a substitute for chemical pesticides ultimately reducing the environmental impact [1]. However, research studies report that *B. thuringiensis* can impact non-target organisms, such as beneficial insects and soil microbiota as well as pose ecological risks due to horizontal gene transfer between *B. thuringiensis* crops and wild relatives [2]. Also, *B. thuringiensis* display a similar repertoire of potential virulence genes on the chromosome as the human pathogenic *B. cereus sensu strictu* [3], and these genes can be actively expressed in *B. thuringiensis*, eventually resulting in foodborne disease. Indeed, there is growing evidence that exposure of agricultural staff and spill-over of *B. thuringiensis* into the food chain may pose a risk to human health [4-8]. To address these risks while discussing whether *B. thuringiensis* should be used as biopesticide, we manually selected the following stakeholder groups: Food Safety Authority (FSA), consumer, and farmer. Stakeholders identified the following question for a prospective discussion: “Should *B. thuringiensis* continue to be used as biopesticide in Central Europe?”.

**Case scenario 2: Wild Boar Hunting**

The second case scenario tackled the ongoing debates about hunting in Germany [9-11], specifically focusing on one species, which is wild boar (*Sus scrofa ferus*). Over the past 30 years, wild boar populations in Europe have steadily increased due to factors such as reduced predator presence and changes in agricultural practices [12]. This rise has been linked to significant damage to agricultural crops [13], decreased plant biodiversity [14], and the spread of infectious diseases such as African Swine Fever [15]. Meanwhile, animal protection activists highlight unnecessary animal suffering caused by hunting and propose alternative methods for population control and crop protection [9], such as immunological contraception [16], fencing [17], and dissuasive feeding [18]. To tackle these controversies, hunters, animal protection representatives, and farmers were manually selected as key stakeholder groups. Stakeholders framed the central question as: "Should hunting of wild boars be forbidden in Germany?”. They then specified their positions on whether wild boar hunting should be prohibited in Germany.

**Description and background information of case scenarios used for evaluation**

**Case scenario 3: *Campylobacter spp.***

*Campylobacter (C.) jejuni* and *C. coli* are the most frequently reported bacterial causes of foodborne gastroenteritis in Europe and many industrialized countries. Despite decades of control efforts, human infection rates have continued to rise, making campylobacteriosis the most common zoonotic disease in the EU. In 2019, the EU reported over 220,000 confirmed cases, but this number represents only a fraction of the true incidence. Accounting for underreporting, it is estimated that about nine million human cases occur each year across the EU, with an annual economic burden of approximately €2.4 billion. In Germany, the Robert Koch Institute reported an increase from 22,000 cases in 2006 to nearly 56,000 in 2019, reflecting a growing public health challenge. The majority of infections are sporadic and associated with the handling and consumption of contaminated poultry meat, particularly broiler carcasses. Quantitative source attribution studies consistently indicate that poultry products account for between 50 and 80 percent of human cases.

Human campylobacteriosis is generally self-limiting, manifesting as acute gastroenteritis with diarrhea, abdominal pain, and fever. However, in a small but significant proportion of patients, infection can lead to serious post-infectious complications such as Guillain–Barré syndrome, reactive arthritis, and inflammatory bowel disease. The mechanism underlying these complications, particularly Guillain–Barré syndrome, involves molecular mimicry between bacterial lipooligosaccharides and human nerve gangliosides, triggering autoimmune responses. The disease burden, therefore, extends beyond the initial infection period, with long-term health impacts and associated costs. The infectious dose for *Campylobacter* is very low—typically below 500 cells—meaning that even minimal contamination of meat can cause human illness. As a result, small reductions in bacterial counts on carcasses can lead to disproportionately large public health benefits.

The *Campylobacter* contamination of poultry meat originates primarily from the intestinal colonization of broilers, where bacterial loads can exceed $10^9$ CFU per gram of intestinal content. Birds remain asymptomatic, which makes on-farm detection and control challenging. Transmission occurs through the environment, contaminated water, feed, pests such as flies, and human activity within farms. During slaughter, contamination can increase through cross-contact during scalding, defeathering, and evisceration steps. A colonized flock is about 30 times more likely to result in contaminated carcasses than a non-colonized one. Despite hygiene improvements and the introduction of a process hygiene criterion (PHC) in 2017 under EU Regulation No. 2073/2005—setting a limit of 1,000 CFU/g on 50 carcass samples per batch—no significant reduction in human case rates has been observed. Quantitative microbial risk assessment (QMRA) models have shown that a one-log reduction (tenfold) in contamination levels at slaughter could halve the number of human campylobacteriosis cases attributable to poultry meat.

Economic analyses illustrate the scale of both the health and control costs. The mean cost per case of campylobacteriosis in Europe is estimated at around €267, including medical care and productivity losses. In the Netherlands, the total national burden is about €82 million annually. At the EU level, the implementation of effective control measures that achieve a one-log reduction in contamination could save approximately €1.2 billion per year in avoided healthcare costs and lost productivity. By comparison, the costs of on-farm and processing interventions are estimated at €0.01–0.03 per kilogram of poultry meat, corresponding to a total industry expenditure of roughly €200 million annually—indicating a clear net societal benefit. However, achieving such reductions remains technically difficult, as *Campylobacter* is highly resistant to environmental stress and persists in water films, aerosols, and organic residues. Moreover, the bacterium's widespread presence in poultry flocks and slaughter environments makes complete elimination unrealistic.

Intervention strategies along the poultry production chain include improved biosecurity, optimized cleaning and disinfection, pest control, and the avoidance of flock thinning. On-farm measures reduce flock colonization risk but cannot guarantee *Campylobacter*-free production. During processing, the use of scalding temperature control, logistic slaughter, and carcass rinsing with approved antimicrobials can achieve limited reductions. Innovative technologies such as cold plasma, UV light, pulsed electric fields, and bacteriophage application have demonstrated 1–3 log reductions in experimental settings. Modified atmosphere packaging can further inhibit bacterial survival during storage and distribution. Despite their promise, these interventions vary in cost, scalability, and consumer acceptance. Consequently, the debate has shifted toward whether microbiological criteria for food safety should be tightened to incentivize adoption of effective control measures throughout the chain.

A key question in the current EU context is whether replacing the existing process hygiene criterion with a legally binding food safety criterion (FSC) for *Campylobacter* in broiler meat would meaningfully reduce public health risks. A food safety criterion would define compliance for the end product, ensuring that only meat meeting microbiological standards could be marketed. Modeling suggests that stricter limits could prevent tens of thousands of human cases annually. However, enforcing such a criterion could have major implications for producers, including higher testing costs, product reprocessing, or batch rejection if limits are exceeded. These actions could in turn generate significant food waste and economic loss, particularly for small and medium-sized enterprises. The challenge is to design a system that maximizes health benefits while maintaining proportionality and minimizing unintended consequences such as resource inefficiency and food disposal.

From a One Health and sustainability perspective, *Campylobacter* control must integrate public health protection, animal production realities, and environmental goals. The European Green Deal and the Farm-to-Fork Strategy call for halving food waste by 2030 and improving resource efficiency across the food chain. Stricter safety standards that result in extensive product waste would conflict with these objectives. Therefore, modern risk-based approaches emphasize dynamic criteria supported by predictive modeling, rapid diagnostics, and smart sampling strategies. These tools allow for earlier intervention, targeted corrective actions, and better alignment between food safety and sustainability goals. In this context, achieving the optimal balance between food safety, economic feasibility, and environmental responsibility is not merely a regulatory issue but a broader societal negotiation based on scientific evidence, transparency, and shared responsibility across the food system.

In summary, *Campylobacter* remains the most significant bacterial foodborne pathogen in Europe, with persistent and growing public health and economic impacts despite long-standing control efforts. The bacterium's ubiquity in poultry production, its resilience within the food chain, and the low infectious dose make complete eradication infeasible. Quantitative evidence shows that even moderate reductions in carcass contamination can substantially decrease human illness, supporting continued investment in risk-based control strategies. However, any regulatory tightening—such as shifting from a process hygiene criterion to a food safety criterion—must be guided by proportionality, feasibility, and sustainability principles. Balancing food safety improvements with economic realities and environmental goals will require integrated surveillance, innovation along the production chain, and harmonized implementation of One Health approaches. The forthcoming negotiation will therefore focus on determining how these objectives can be aligned to achieve maximum public health protection without compromising food security, industry viability, or the EU's sustainability commitments.

## Case scenario 4: Oak processionary moth

The oak processionary moth (*Thaumetopoea processionea*, OPM) has become a growing concern across Central Europe over the past two decades. Originally regarded as a rare phenomenon confined to warm, lowland oak forests, this thermophilic species has expanded its range northward and westward, largely in response to milder winters and prolonged dry summers. The moth's larvae feed on oak foliage, and repeated defoliations can weaken trees, reduce growth, and make them more vulnerable to secondary pests and diseases. However, the problem extends well beyond forestry and ecology. From a broader One Health perspective, OPM exemplifies a complex, multi-dimensional risk at the intersection of environmental, human, and animal health[19,20].

The central challenge arises from the caterpillar's urticating hairs, which contain the toxin thaumetopoein and can cause severe irritations upon contact with skin, eyes, or respiratory mucosa. These hairs are barbed, lightweight, and can be dispersed by wind, remaining active in the environment for years even after the larvae have pupated. Human exposure may occur directly through outdoor activities or indirectly via contaminated materials. Documented effects include dermatitis, conjunctivitis, and in rare cases, respiratory distress or allergic reactions[21,22]. Outbreaks have been reported to affect thousands of people during peak infestation years. Healthcare systems, local administrations, and occupational sectors such as forestry and landscaping have all experienced an increased burden due to the medical and logistical consequences of large-scale OPM infestations[19].

In France, Germany, the Netherlands, and other affected countries, epidemiological data confirm that the OPM is no longer an occasional nuisance but a recurring seasonal health hazard. A recent expert report by the French Agency for Food, Environmental and Occupational Health and Safety[19] classified the processionary moths—both of oak and pine—as a significant public health concern. The report emphasizes that exposure to these caterpillars is expected to increase as climate change alters temperature patterns and extends the species' active season. ANSES also warns of potential underreporting, as mild cases of skin irritation or ocular inflammation are often not formally registered, leading to an incomplete assessment of the true population-level impact. The report further notes that professional exposure, particularly among forestry workers, gardeners, and municipal employees, constitutes a specific risk group, with over 60 percent of surveyed outdoor workers reporting symptoms following contact with processionary caterpillars[19].

At the same time, the ecological implications are becoming increasingly evident. As shown in long-term field studies in Bavaria and other regions of Germany, repeated defoliations can lead to the gradual decline of oak vitality. While single defoliation events are seldom fatal for mature trees, consecutive outbreaks reduce photosynthetic capacity, hinder regrowth, and predispose trees to infestation by secondary pests such as the oak buprestid beetle (*Agrilus biguttatus*) or fungal pathogens like *Armillaria*. These cascading effects can ultimately contribute to local dieback and the degradation of mixed oak ecosystems[20]. The loss of canopy cover also has implications for biodiversity, carbon sequestration, and forest recreation, making OPM not only a biological pest but also an economic and social problem[19].

Efforts to control OPM populations face significant trade-offs. Mechanical removal of nests and the use of biological insecticides such as *Bacillus thuringiensis* (*Bti*) have proven effective under certain conditions, yet both methods have limitations. Mechanical removal is labor-intensive and poses health risks for workers, while biological control requires precise timing and careful environmental management to avoid non-target effects. Large-scale aerial spraying has been largely abandoned in Europe due to its potential ecological consequences and low public acceptance[19]. As a result, integrated management strategies now rely on a combination of early detection, targeted interventions, and public risk communication. However, as pointed out by both national and regional agencies, there are still no harmonized thresholds to define when the health or ecological risk of OPM requires active intervention[20]. This lack of standardized criteria often leaves local authorities to decide on a case-by-case basis, balancing immediate public health concerns against long-term environmental and financial costs.

The oak processionary moth thus represents a contemporary case of risk governance under uncertainty, where scientific knowledge, environmental change, and societal expectations intersect. The issue highlights the need for coordinated, interdisciplinary management approaches that integrate ecological monitoring, medical preparedness, and public communication. It also raises fundamental questions about how to prioritize among competing objectives: protecting human health, maintaining forest ecosystem integrity, and ensuring social and economic sustainability. As OPM continues to expand its range and the frequency of outbreaks increases, decision-makers across Europe are faced with the urgent task of developing coherent, science-based frameworks for prevention and control that are acceptable to all sectors of society[19].

In summary, the OPM problem encapsulates the broader challenges of managing emerging biological and environmental risks in a changing climate. It demands collaboration between health, environmental, and forestry sectors, grounded in evidence and informed by public trust. The forthcoming risk negotiation exercise will therefore use the OPM case as an example to explore how different types of expertise and evidence can be brought together to support balanced, transparent, and resilient decision-making under complex, real-world conditions.

**Supplementary Text S2**

**Detailed description of the individual steps of the AI-assisted negotiation-centered risk analysis process**

*Step 1: Setting the rules and stakeholder roundtable establishment (stakeholder selection)*

Once a problem for negotiation is identified, the negotiation rules must be clearly defined to ensure transparency and minimize misunderstandings or potential conflicts during the process. This includes specifying the modalities and organization of the meetings, outlining the roles of participating stakeholders - such as if someone holds veto rights or initiates the negotiation - and detailing the process for progressing through each step. It should also be specified if and how participants may revisit previous steps, as well as the conditions for reaching consensus, such as by majority vote or full agreement. It is also possible to determine through negotiation simulations if certain rules are ineffective. Following these initial steps, and based on a discussed area, the selection of relevant stakeholders is done. The selection process should facilitate knowledge sharing and exchanging perspectives as these factors influence the identification of trade-offs during the risk assessment, management, and communication process [23]. For our purposes, a stakeholder is defined as any individual, group, or organization with a vested interest in a discussed area. Stakeholder groups are selected within public, private, or civic sectors to reflect various interests. To identify relevant stakeholders, publicly available information, e.g. peer-reviewed articles from scientific literature archives such as PubMed Central®, may be used. One can use LLMs such as the latest models from OpenAI© [24] or Alphabet© [25], or open-source alternatives (such as Hugging Face [26], Llama [27], or Nvidia [28]), to suggest relevant stakeholder groups for a given problem (see step 2). After identifying relevant stakeholder groups and their representatives, their roles in the negotiation process, such as veto holders, moderator and other responsibilities need to be determined.

*Step 2: Problem formulation*

A clear problem formulation is necessary before negotiation begins. This can range from a broad, question-based format to a more detailed problem description or a list of problems to be discussed within the previously outlined discussion area. It is essential to ensure that all stakeholders can fully engage with this process. Initially, LLMs can offer a suggested problem definition and subsequently aid in refining the problem formulation. For analysis of confidential information, a user can also switch to the usage of open-source models [26-28], or models within local environments [29]. Stakeholder representatives must refine and agree upon a more precise definition, following a human-in-the-loop (HIL) approach. When refining the problem definition, it is important to focus on a single issue while capturing its full complexity across all dimensions.

*Step 3: Risk assessment and valuation*

At this step, stakeholders aim to agree on the list of assessed risks/issues, evaluate possible risk management options, and weigh the advantages and disadvantages of each option. The process ultimately leads to determining the risk-based value of each option. This part of the approach, initially detailed by Ehling-Schulz *et al.* [23], experienced the most substantial further development and was subdivided into five concrete steps (3a-3e) to be followed during a first exercise as follows: (3a) collect negotiation issues, i.e. effect dimensions; (3b) find consensus on effect dimensions; (3c) collect risk management options for each of the effect dimensions; (3d) agree on collected options; and (3e) individually value each option based on their importance. Practically, each stakeholder is to be instructed to specify their position in relation to the negotiated question by providing information in the form of plain text, potentially within predefined categories such as description, short-term goals, general aim, scientific foundation, ethical foundation. This information will then be used to prompt LLM for the simulation of the preliminary negotiation between the stakeholders with the aim of collecting possible issues (i.e. effect dimensions) and risk management options (i.e. prospective solutions). Thus, stakeholders will end up with a list of concrete issues and a fixed number of options per issue to be further discussed. Next, stakeholders are asked to quantify their preferences by assigning a score to each available option. Using a pre-defined number of points (e.g. 100), each stakeholder distributes their scores across the identified effect dimensions (i.e. issues) according to their importance, with higher scores indicating higher priority. Once all issues are scored, the specific options for each effect dimension are then also scored based on their importance. Each stakeholder also needs to define a minimum threshold for an agreement. These scoring principles allow representatives to prioritize risk negotiation issues of greatest concern, while also highlighting areas where compromises may be possible. The individual scores are kept confidential and will be used in LLM-assisted negotiation simulations as described in step 4 [30].

*Step 4: Risk Negotiation*

The risk negotiation step aims to find a compromise (‘Nash’ equilibrium) between the stakeholder preferences outlined and scored in step 3. This compromise-based risk management approach, referred to as a "deal," consists of a set of proposed risk management solutions (options) designed to address the various identified issues. For example, in one of our real case scenarios, a key question concerned the use of a biopesticide. One issue, "Consumer Safety and Health Risk" (Issue A), involved options based on the level of safety measures to be applied for this biopesticide: "strict," "moderate," or "minimal" (Options A1, A2, and A3, respectively). Ultimately, multiple issues (A-F) were identified, each linked to three potential options (e.g., A1-A3, B1-B3, etc.). The resulting deal from the risk negotiation process is then a selected combination of options addressing each issue (e.g., A1, B2, C3, D3, E1, F2). Since stakeholders assigned scores to each option based on their preferences (step 3), the deal acceptance criteria can be understood as reaching a specific numerical threshold, and the negotiation process is characterized as a non-zero-sum game. In brief, the risk negotiation is performed with LLM-assisted negotiation simulation within a multi-agent framework described by Abdelnabi *et al.* [30]. In our approach, agents represent the stakeholders and are guided by initial prompts containing shared project details, stakeholder-specific confidential scores, and game instructions. Negotiations are launched with a suggestion from a moderator (chosen from one of the involved parties) and proceed over multiple rounds, with agents exchanging the information in plain text form, either supporting previous deals or proposing new ones based on the interaction history. After all rounds, a final deal is proposed by the moderator. The multiple iterations of the LLM-based simulation are conducted to ensure the reproducibility of the results and to account for fluctuations due to the different initial conditions, i.e. moderator role (see Fig. S2 and Fig. S4 for practical examples of the moderator role influence). The ultimate result is a distribution of suggested deals. Obtained deals are to be discussed among the stakeholders to either agree on one of them or modify the deals to reach a consensus. If necessary, stakeholders can check the history of each of the deals proposed during the negotiation simulation, track how the proposals were formulated, and identify what was the rational behind (see an example of such an analysis in case scenario 1 below; also, Movie S1, provided as Supplementary Material, exemplarily illustrates a potential analysis pipeline).

**Supplementary Text S3**

**Template Scoring Case scenario 1 (*B. thuringiensis*)**

**Step 3d: Valuation – Scoring options of the stakeholders.**

Stakeholder *(please check the respective box)*

☐ Consumer
☐ Farmer
☐ Food Safety Authority

Date of scoring:
Quick guide for the scoring scheme

| |
|---|
| *1. General rules*<br>You have **100 points** that can be used to define your preferences.<br>The higher the value, the higher the preference.<br>You do not know the scores of the other parties, but you do know the descriptions of the other parties, so you can share your guesses on what other may think (*optional*).<br>Your goal is to define scores reflecting your preferences, but you may also show possible areas where a comprise is feasible.<br><br>*2. Follow the 2-step scoring approach:*<br>First, for each issue define the maximum score (points) you put on this issue as follows:<br>• Max score weights the importance of the issue to you<br>• You can put scores from 0 to 100:<br>  o 0- the lowest importance<br>  o 100- the highest importance<br>• sum of scores of all issues should be 100 in total (remember: you have 100 points to be used – general rule!)<br><br>Secondly, for each suboption of the issues define the option scores as follows:<br>• Option scores define the preferences and your flexibility<br>• option scores may not be higher than the max score of this issue (defined at 1)<br>• at least one of the suboptions (the one reflecting your preferences the most) should be scored equally to the max score chosen for this issue<br>• the other suboptions may be scored according to your preferences using scores from 0 to the max score chosen for this issue<br>• Limit yourself to natural numbers |

*Now, please provide your scorings for the following issues and options:*

| ***Issue and suboptions*** | ***Your score*** |
|---|---|
| ***A:*** *Consumer Safety and Health Risk*<br>*This issue pertains to the potential health risks associated with the use of Bt as a biopesticide.* | *(max score)* |
| A1: "Strict Safety Measures": Implement rigorous testing and regular monitoring to ensure Bt does not pose any health risks to consumers until Bt biopesticide application have been proven to be harmless for humans | |
| A2 "Moderate Safety Measures": Conduct regular assessments and testing but with less stringent monitoring. Implementing a waiting time. | |
| A3 "Minimal Safety Measures": Rely on existing safety data with minimal additional testing and monitoring until adverse health effects have been demonstrated. | |
| *Optional: General comment on the scoring for Issue A*<br>________________________________________ | |

| ***Issue and suboptions*** | ***Your score*** |
|---|---|
| ***B:*** *Environmental Impact and Biodiversity*<br>*This issue concerns the long-term environmental impact of Bt on biodiversity.* | *(max score)* |
| B1 "Intensive Monitoring": Conduct extensive studies and continuous monitoring of Bt's impact on non-target organisms and ecosystems until Bt biopesticide application have been proven to be harmless for the environment and the biodiversity. | |
| B2 "Balanced Monitoring": Implement moderate monitoring efforts to assess the environmental impact periodically. | |
| B3 "Limited Monitoring": Rely on existing environmental impact studies with minimal new monitoring efforts until adverse environmental effects have been demonstrated. | |
| *Optional: General comment on the scoring for Issue B*<br>________________________________________ | |

| ***Issue and suboptions*** | ***Your score*** |
|---|---|
| ***C:*** *Regulation and Monitoring*<br>*This issue addresses the regulation and oversight required for the use of Bt.* | *(max score)* |
| C1 "Stringent Regulation and Monitoring": Establish strict regulations, including genome sequencing and tight control over Bt strains used. | |
| C2 "Moderate Regulation and Regular Monitoring": Implement standard regulations with periodic reviews and monitoring. | |
| C3 "Minimal Regulation and Basic Monitoring": Maintain current regulatory practices with minimal additional oversight. | |
| *Optional: General comment on the scoring for Issue C*<br>________________________________________ | |

| ***Issue and suboptions*** | ***Your score*** |
|---|---|
| ***D:*** *Resistance Development*<br>*This issue focuses on the potential for pests to develop resistance to Bt and the strategies to manage it.* | *(max score)* |
| D1 "Integrated Pest Management (IPM)": Promote IPM strategies including crop rotation and biological controls alongside Bt use. | |
| D2 "Regular Monitoring and Adaptation": Monitor pest resistance levels and adapt Bt usage accordingly. | |
| D3 "Status Quo": Continue current Bt usage practices without additional resistance management strategies. | |
| *Optional: General comment on the scoring for Issue D* | |

| ***Issue and suboptions*** | ***Your score*** |
|---|---|
| ***E:*** *Economic Viability and Sustainability*<br>This issue concerns the economic impact of Bt use on farming and its sustainability. | *(max score)* |
| E1 "High Support for Farmers": Provide significant support and subsidies for farmers using Bt to ensure economic viability and sustainability. | |
| E2 "Moderate Support for Farmers": Offer moderate financial support and incentives for Bt use. | |
| E3 "Minimal Support for Farmers": Rely on the current market dynamics with minimal additional support for farmers using Bt. | |
| *Optional: General comment on the scoring for Issue E* | |

| ***Issue and suboptions*** | ***Your score*** |
|---|---|
| *F: Communications*<br>*This issue addresses the communication required for the use of Bt.* | *(max score)* |
| F1 "Comprehensive Communication": Ensure comprehensive communication with the public about the safety measures and labelling of sprayed products. | |
| F2 "Regular Communication": Provide regular updates to the public about the safety and usage of Bt. | |
| F3 "Minimal Communication": Offer basic communication to the public about Bt use. | |
| *Optional: General comment on the scoring for Issue F* | |

*Optional: General comments on the scoring*

**Supplementary Text S4**

**Template Scoring Case scenario 2 (Wild Boar Hunting)**

Stakeholder *(please check the respective box)*

☐ Hunter
☐ Farmer
☐ Animal protection representative

Date of scoring:
Quick guide for the scoring scheme

| |
|---|
| *1. General rules*<br>You have **100 points** that can be used to define your preferences.<br>The higher the value, the higher the preference.<br>You do not know the scores of the other parties, but you do know the descriptions of the other parties, so you can share your guesses on what other may think (*optional*).<br>Your goal is to define scores reflecting your preferences, but you may also show possible areas where a comprise is feasible.<br><br>*2. Follow the 2-step scoring approach:*<br>First, for each issue define the maximum score (points) you put on this issue as follows:<br><br>• Max score weights the importance of the issue to you<br>• You can put scores from 0 to 100:<br>    o 0- the lowest importance<br>    o 100- the highest importance<br>• sum of scores of all issues should be 100 in total (remember: you have 100 points to be used – general rule!)<br><br>Secondly, for each suboption of the issues define the option scores as follows:<br><br>• Option scores define the preferences and your flexibility<br>• option scores may not be higher than the max score of this issue (defined at 1)<br>• at least one of the suboptions (the one reflecting your preferences the most) should be scored equally to the max score chosen for this issue<br>• the other suboptions may be scored according to your preferences using scores from 0 to the max score chosen for this issue<br>• Limit yourself to natural numbers |

*Now, please provide your scorings for the following issues and options:*

| ***Issue and suboptions*** | ***Your score*** |
|---|---|
| ***A: Wild boar population control***<br>*This issue focuses on the management of wild boar populations in Germany, taking into account population control, disease control, socio-economic impacts, and ethical considerations.* | *(max score)* |
| A1 "Strict Control": Implement strict control measures, including regulated hunting seasons and quotas to maintain a balanced population and prevent overpopulation and disease spreading. | |
| A2 "Moderate Control": Allow moderate hunting with specific guidelines to manage the population without overhunting and considering disease spreading. | |
| A3 "Minimal Control": Implement minimal control measures, focusing on non-lethal methods of population control such as relocation and sterilization programs and considering disease spreading. | |
| *Optional: General comment on the scoring for Issue A* ________________________________ | |

| ***Issue and suboptions*** | ***Your score*** |
|---|---|
| ***B: Promoting Coexistence and Ethical Treatment*** *This issue highlights the importance of promoting coexistence with wildlife and ensuring the ethical treatment of wild boars in population management.* | *(max score)* |
| B1 "Coexistence Priority": Prioritize coexistence measures, including extensive use of non-lethal methods and public education on living with wild boars. | |
| B2 "Balanced Approach": Combine ethical hunting practices with non-lethal methods to promote coexistence and ethical treatment. | |
| B3 "Hunting Priority": Focus primarily on ethical hunting as a means of population control, with minimal emphasis on non-lethal methods. | |
| B4 "No Tourism Priority": Prioritize coexistence measures, including extensive use of non-lethal methods and public education on living with wild boars, strictly forbidding hunting tourism. | |
| *Optional: General comment on the scoring for Issue B* ________________________________ | |

| ***Issue and suboptions*** | ***Your score*** |
|---|---|
| ***C**: **Economic Viability and Agricultural Protection*** *This issue focuses on the economic viability of farming communities and the protection of agriculture from wild boar damage.* | *(max score)* |
| C1 "High Compensation": Provide high compensation to farmers for crop damage or livestock disease management options caused by wild boars, funded by hunting licenses and related tourism. | |
| C2 "Moderate Compensation": Offer moderate compensation to farmers for crop damage or livestock disease management options and implement preventive measures such as fencing and repellents. | |
| C3 "Low Compensation": Provide low compensation to farmers for crop damage or livestock disease management options with a focus on self-sustaining preventive measures by farmers. | |
| *Optional: General comment on the scoring for Issue C* ________________________________ | |

| ***Issue and suboptions*** | ***Your score*** |
|---|---|
| ***D: Cultural Significance and Public Awareness*** *This issue addresses the cultural significance of hunting, public awareness, and education on wild boar management practices.* | *(max score)* |
| D1 "Promote Hunting Culture": Strongly promote the cultural heritage and conservatory role of hunting through public campaigns and education. | |
| D2 "Balanced Cultural Promotion": Promote both hunting culture and conservatory role of hunting and the importance of ethical wildlife management equally. | |
| D3 "Minimal Hunting Promotion": Focus on promoting non-hunting aspects and education of wild boar management and cultural significance. | |
| *Optional: General comment on the scoring for Issue D* | |

| ***Issue and suboptions*** | ***Your score*** |
|---|---|
| ***E: Scientific Data Generation and Monitoring*** *This issue addresses the generation of scientific data and monitoring of wild boar populations including crop damage, occurrence of diseases transferred by wild boars, and genetic population structure.* | *(max score)* |
| E1 "Comprehensive Monitoring": Implement comprehensive monitoring programs, including advanced tracking and data collection methods, including animal welfare issues. | |
| E2 "Moderate Monitoring": Establish moderate monitoring efforts with regular data collection and reporting, including animal welfare issues. | |
| E3 "Basic Monitoring": Establish basic monitoring activities with minimal resource allocation. | |
| *Optional: General comment on the scoring for Issue E* | |

*Optional: General comments on the scoring*

**Supplementary Text S5**

**Final Issue-Option formulations Case scenario 3:** ***Campylobacter spp.***

Issue A: "Type and Stringency of Campylobacter Microbiological Criterion (PHC vs. FSC)"

A1. "Status quo PHC": Maintain the existing PHC only (no FSC). Continue relying on process controls and corrective actions without imposing batch-based market prohibitions.
A2. "Phased FSC (lenient limit)": Introduce an FSC with a relatively high limit and c>0 acceptance, phased in gradually with risk-based sampling. This would ease adaptation for SMEs and reduce waste.
A3. "Risk-based FSC (moderate limit)": Establish an FSC aligned to public health targets, with mandatory batch actions when limits are exceeded. Apply risk-based sampling frequency and implement targeted waste mitigation measures.
A4. "Strict FSC (low limit & immediate enforcement)": Adopt a stringent FSC with a low limit and immediate withdrawal or recall of non-compliant batches to maximize public health protection.
=================

Issue B: "Primary control focus along the supply chain"

B1. "Processing & Retail Focus": Concentrate efforts on enhancing hygiene practices and implementing advanced decontamination technologies at processing facilities, while also improving retail standards and transparency to ensure product safety. This approach aims to reduce contamination in the supply chain and provide consumers with clear information on poultry safety, without focusing primarily on on-farm interventions.
B2. "Process-first": Prioritize slaughterhouse and processing controls, including decontamination technologies, as the main points for risk reduction.
B3. "Balanced multi-hurdle": Distribute control measures and investment evenly across the farm, processing, retail, and consumer stages.
B4. "Consumer & transparency focus": Center efforts on labeling, behavior change, and transparency initiatives to reduce household-level contamination risks.
B5. "Environment-first": Emphasize control of environmental leakage (e.g., runoff, vectors, VBNC survival) and enforce tighter environmental safeguards.
=================

Issue C: "Cost burden sharing and food-waste mitigation mechanisms"

C1. "Producer liability": Industry bears full compliance and non-compliance costs. Failed batches follow standard disposal routes without subsidies. Additional costs cannot be passed to the consumer
C2. "Targeted public co-funding": Provide temporary public financial support (subsidies or tax credits) for validated control measures and SME adaptation, to prevent consumer price increases.
C3. "Regulated pass-through with transparency": Permit cost pass-through to consumers, conditional on transparent reporting and performance benchmarking by retailers. There will be no subsidies or tax credits
C4. "Waste-minimization toolkit": Fund innovations such as dynamic shelf-life management, reprocessing or valorization options, and case-by-case batch risk assessment to reduce avoidable waste.
C5. "Balanced cost-sharing compact": Combine producer obligations with capped public support and retailer co-investment, tied to measurable public health outcomes.
=================

Issue D: "Implementation scope: uniform EU rule vs tailored, risk-based flexibility"

D1. "Uniform EU standard": Apply one EU-wide threshold and sampling plan, with minimal derogations, for simplicity and fairness.
D2. "Harmonized baseline with upward flexibility": Define a common minimum baseline for PHC/FSC, allowing Member States or retailers to implement stricter measures if desired.
D3. "Risk-tailored application": Adapt criteria and sampling based on regional risk, flock characteristics, or processor capacity. Allow case-by-case adjustments and SME relief mechanisms.
D4. "Voluntary adoption & transparency": Avoid binding EU-wide rules; instead, rely on voluntary schemes supported by mandatory public reporting to enable consumer-driven accountability.
=================

Issue E: "Implementation scope: Environmental foodprint of a new foodsafety criteria"

E1. "Do not intervene": Click to edit option description
E2. "Intervene with the guideline to reach CO2 equivalent": Click to edit option description
E3. "Intervene with the legally binding text to reach CO2 equivalent": Click to edit option description
=================

**Supplementary Text S6**

**Final Issue-Option formulations Case scenario 4: OPM**

Issue A: "Control Method Mix & Intensity"

A1. "Broad Chemical Suppression": Authorize area-wide insecticide applications, including non-selective actives, during the larval stages to quickly reduce populations. Ecological safeguards would be secondary to urgent public health protection.
A2. "Targeted Biocides (Bt/Diflubenzuron) within IPM": Permit tightly targeted applications (timed to early larval stages, using drift-minimizing techniques) near high-use sites, as part of an Integrated Pest Management (IPM) plan. Require mechanical nest removal and worker protection measures.
A3. "Mechanical First, Biocides as Back-up": Prioritize mechanical nest removal, vacuuming, and hot-spot sanitation. Use biocides only when mechanical control is not feasible or where health risks remain high.
A4. "Nature-Based Preference": Emphasize predator habitat enhancement, diversified plantings, and trials with biological control agents. Restrict chemical or biocidal interventions to exceptional circumstances with strict non-target protection thresholds.
A5. "No Sprays / Monitor & Tolerate": Suspend all chemical and biocidal spraying. Manage public exposure through site closures, advisories, and personal protection while allowing ecosystems to self-regulate naturally.
=================

Issue B: "Targeting, Prioritization & Timing of Interventions"

B1. "Blanket Coverage": Treat all infested oak stands within an administrative area during the season, irrespective of public use intensity, to simplify operations and provide reassurance to the public.
B2. "Model-Driven, City/Region-Wide Targeting": Use predictive tools (phenology and risk maps) to prioritize treatment windows and clusters across the jurisdiction. Sequence interventions for maximum impact with minimal collateral effects.
B3. "High-Use Sites First": Focus on high-risk locations such as schools, playgrounds, parks, hospitals, and commuter corridors. Elsewhere, rely on advisories and monitoring rather than active control.
B4. "Reactive Hot-Spot Response": Act only after confirmed public complaints or health incidents. Maintain rapid-response teams for nest removal and communication efforts.
B5. "Access Management & Public Guidance Only": Rely primarily on closures, signage, and exposure forecasts rather than direct suppression. Intervene only during exceptional outbreaks.
=================

Issue C: "Decision-Making Framework & Public Communication"

C1. "Health Authority Leadership": Public health authorities lead the decision-making process, ensuring legal compliance and prioritizing urgent health measures while consulting stakeholders.
C2. "Municipal Coordination": Municipalities coordinate with standard operational procedures to ensure compliance, involving stakeholder collaboration for comprehensive intervention strategies.
C3. "Collaborative Task Force": A task force comprising health, legal, and community/stakeholder representatives to ensure decisions are balanced, transparent, and legally sound.
C4. "Community-Led Engagement": Community groups play a significant role in shaping interventions, ensuring public concerns are addressed and communicated effectively.
C5. "Minimalist Approach with Legal Notices": Authorities issue legal notices and advisories with minimal public engagement, focusing on compliance and operational efficiency.
=================

Issue D: "Funding & Cost-Sharing"

D1. "State/National Co-Financing": Provide higher-level grants or subsidies to cover a substantial share of municipal and forestry costs, contingent on risk-based plans and transparent reporting.
D2. "Municipal Primary, Targeted Top-Ups": Assign core funding to municipalities, with support from regional or health authorities for priority areas like schools and public parks.
D3. "Landowner / 'Beneficiary Pays'": Require both private and public landowners to finance control measures on their properties, reserving public funds for high-risk communal areas.
D4. "Levy & Insurance Mix": Introduce a small local levy and promote insurance or contractor pooling schemes. Revenues would be ring-fenced for prevention and rapid response programs.
D5. "Minimal Public Expenditure": Restrict public funding to communication and awareness activities. Individuals and landowners bear the primary cost of prevention and intervention.
D6. "Collaborative Funding Consortium": Establish a multi-stakeholder consortium that pools financial resources from municipalities, regional governments, landowners, and private entities to collectively finance OPM management. This collaborative model ensures shared responsibility and risk, with contributions scaled based on the benefit received, land size, and infestation severity.
=================

Issue E: "Ecological Safeguards & Non-Target Protection"

E1. "Strict Safeguards": Prohibit spraying near water bodies and during pollinator activity. Enforce low-drift technology, buffer zones, mandatory monitoring, and public disclosure of results.
E2. "Enhanced Safeguards": Permit targeted biocide use only with drift-reduction equipment and weather/time-of-day constraints. Apply exclusions for sensitive habitats.
E3. "Balanced Safeguards": Follow standard national environmental guidelines for non-target protection, prioritizing public health interventions in high-risk areas.
E4. "Pragmatic Minimal Safeguards": Adhere to basic legal environmental requirements; apply further restrictions only if monitoring shows measurable harm to non-target species.
E5. "Case-by-Case Exemptions": Allow temporary exemptions from safeguards during severe public health emergencies, followed by post-event ecological audits.
=================

Issue F: "Proactive OPM Management Strategies"

F1. "Comprehensive Monitoring Systems": Implement an integrated monitoring system using technology, citizen science, and professional surveys to detect OPM early in public and private spaces. This approach enables timely interventions and data-driven decision-making.
F2. "Community Education and Engagement": Develop and disseminate educational programs for residents, schools, and local businesses about OPM identification, safe interaction practices, and reporting protocols. Encourage public participation in monitoring and reporting efforts to enhance community resilience.
F3. "Ecological Habitat Management": Promote ecological strategies such as diversifying plant species in urban areas and enhancing habitats for natural OPM predators. These measures aim to create a balanced ecosystem that naturally regulates OPM populations, reducing the need for chemical interventions.
=================

**Supplementary Text S7**

**Case scenario 1 *Bti*: Full details of negotiation outcomes**

**Step 4: Modelling results negotiation.**

- the most approved (39.7%) combination of suboptions for the issues by the LLM agents during negotiation was: A1 B2 C2 D2 E1 F2
- Farmers and Food Safety Authorities (FSA) representatives accepted the most approved combination
- The consumer representative didn't accept the most approved combination. Source of disagreement: issue F. The consumer representative would not accept a solution less or equal to F2
- It was decided to disclose the individual scoring of the stakeholders given for issue F:
    - Farmers representative: F1:0 F2:1 F3:3
    - FSA representative: F1:1 F2:9 F3:1
    - Consumers representative: F1:20 F2:0 F3:0

  Background information on the scores: the issue is not top priority for farmers, FSA would accept options higher or equal to F2, consumers rated issue F as second issue of concern for them and would only accept a F1 option.
- Agreement was found to adopt the F1 control option
- Final agreement on the following suboption combination: **A1 B2 C2 D2 E1 F1**
- by achieving this consensus on the suboptions the negotiation of the Bt question is officially closed. Further steps according to the risk negotiation workflow are to follow.

**Supplementary Text S8**

**Case scenario 2 Wild Boar Hunting: Full details of negotiation outcomes**

**Step 4: Modelling results negotiation.**

- the most approved combinations (16.7%, each) of suboptions for the issues by the LLM agents during negotiation were:
    - <u>A2 B2 C2 D2 E1</u>
    - <u>A2 B4 C2 D3 E2</u>
- Farmers and Hunters representatives accepted the two most approved combinations
- The animal protection representative didn't accept the most approved combination. Source of disagreement: issues A and B
- It was decided to partly disclose the priorities of the stakeholders given for the issues
    - Farmers representative: issue C was given highest priority, followed by issue A
    - Hunters' representative: issue A was given highest priority, followed by issue C
    - Animal protection representative: issue B was given highest priority, followed by issue A; issues A+B in combination account for 80% of all sum scores
- As for issue A the hunters' representative could not accept any option less than A2 (moderate)
- However, the animal protection representative shows flexibility in accepting A2 if suboption B4 would be chosen
- Also, the farmers representative shows willingness to accept B4
- Alternatively, the farmers representative shows willingness to accept A3 but in combination with C2, only.
- Agreement was found to negotiate a "suboption package" regarding issues A and B
- less disagreement was observed for issues C, D, E - all stakeholders could live with the proposed agreement
- Final agreement on the following suboption combination was then achieved as follows:
    - **A2 - B4 - C2 - D2 - E1/2**
- by achieving this consensus on the suboptions the negotiation of the hunting question is officially closed. Further steps according to the risk negotiation workflow are to follow.

**Supplementary Text S9**

**Case scenario 3 *Campylobacter spp.:* Full details of negotiation outcomes**

The stakeholders reviewed the most frequently proposed deals (Fig. S6) and provided the following feedback:

- Food Safety Authorities expressed satisfaction with deals **A3-B2-C4-D2-E2** and **A3-B2-C1-D2-E2** (Deal 1). They also indicated a preference for **B2** and strong disagreement with **A2**.

- Industry representatives rejected **A3-B2-C1-D2-E2** because of **C1** and instead suggested **A3-B2-C4-D2-E1** or **A3-B3-C2-D2-E1**. They also indicated a preference for **E1**.

- Consumers accepted **A3-B2-C1-D2-E2**, **A3-B2-C4-D2-E2**, and **A3-B2-C4-D2-E1**, while additionally noting that **B2** was not preferred.

- NGOs accepted **A3-B2-C1-D2-E2**, but expressed a preference for **A2-B3-C4-D2-E2** and indicated that **E1** was not preferred.

During the subsequent exchange, which focused on individual issues and options (Fig. S7), **D2** was accepted on the condition that the final formulation explicitly stated that NGOs would be included in the negotiations. **B3** was accepted, although consumers made their acceptance conditional on **E2**. **E2** was also accepted, with industry representatives proposing the addition that there was a "need to cumulatively develop the guideline." **A3** and **C4** were accepted by all stakeholders, although **A3** was not the preferred option for NGOs.

The final agreed solution was **A3-B3-C4-D2-E2**.

**Supplementary Text S10**

**Case scenario 4 OPM: Full details of negotiation outcomes**

The stakeholders reviewed the most frequently proposed deals (Fig. S8) and provided the following feedback:

- Overall great agreement among stakeholders on issues
- Some stakeholders opted for additional issues
    - Learning from existing experience in areas of origin of OPM and considering uncertainty
    - Pro-active interventions: preparedness and early warning
- Some stakeholders opted for slight modification of issues
- Also, options were partly adapted

During the subsequent exchange, which focused on individual issues and options (Fig. S9), stakeholders agreed on the following solution: **A3-B2-C3-D1-E2-F1-3**

- But adjusting issue F in a sense, that all options are merged into F

**Supplementary Text S11**

**Moderator instructions during evaluation workshop in October 2025**

**Most important**

- Come prepared:
    - read Case specific background paper and stakeholder descriptions carefully
    - Familiarize yourself with stakeholder groups involved and participants allocation (who has which role, which background expertise do they bring)
- Be friendly and polite, but also strict
- Keep strict time management (assisted by facilitator)
- Stick to the agenda - final version here
- Have final proposal of issues-options per case-study ready in a docx file (to be shared with participants via email)

**Roadmap**

- Introduce into topic
- Explain roles and rules properly
- Acknowledge stakeholder descriptions and summary paper on topic as reference for negotiation
- Allow stakeholders to familiarize with stakeholder descriptions ‘10

- **Issue-option voting:**
    - Share proposed issues-options via email
    - Allow stakeholders to view and familarize with proposed issues-options ‘15
    - Briefly emphasize that not each issue might be relevant to all stakeholders given their individual priorities and targets, meaning that one must not agree with each of the issues; options only need to be checked if an issue is of relevance for the stakeholder
    - Start retrieving stakeholder statements one-by-one (‘5 per stakeholder), take notes (useful for re-generation)
    - Start general discussion with clear time limit ’15 min, interact in case individual stakeholders are taking the spotlight
    - Share edited/newly generated issues-options
    - Allow stakeholders to view and familarize with new issues-options ‘10
    - Start voting by calling stakeholders and issues-options one-by-one asking:
        - Do you agree with A, B, C, etc. ?
            - If yes: is there a suitable option you could choose?
                - Yes: ready to go
                - No: how to rephrase/add the option to meet your preferences?*
            - If no: is this issue of relevance for you at all?
                - No: ready to go
                - Yes: how best to rephrase the issue to meet your preferences?*
            - * check with other stakeholders prior to amendments

- **Scoring:**
    - Briefly explain scoring principles and rules and negotiation app functionality
    - Start scoring time ‘25
        - Allow discussion between stakeholders and their observers

- Moderator to guide scoring through monitoring of progress of individual stakeholder, push/guide if necessary

- **Dealing and Consensus finding**:
    - Allow stakeholders to explore individual dashboards and deal charts through the negotiation app ‘10
    - Retrieve stakeholders feedback one-by-one on the TOP5 deals asking:
        - Does one of the proposed deals accomodates you?
            - If yes: which one?
            - If no: why not? Which one comes closest, and what would need to be changed to allow you to agree on that deal?
    - Moderate consensus finding using the deal the majority of stakeholders favors
        - Ask those stakeholders disagreeing how to change to deal
        - If necessary propose to elaborate on rationale behind, also allow for (partly) disclosure of scores, and/or amendments of wording of issues/options
        - Always ask remaining stakeholders if they agree/disagree with proposed changes
    - Propose final deal asking:
        - Is anyone not accepting this deal?
            - If no one replies, the deal has to be accepted
            - If one party replies, start asking why and what might be needed to accept this deal?

**Fig S1. Preliminary Case scenario 1: Combined results – distribution of unique issue-option combinations**

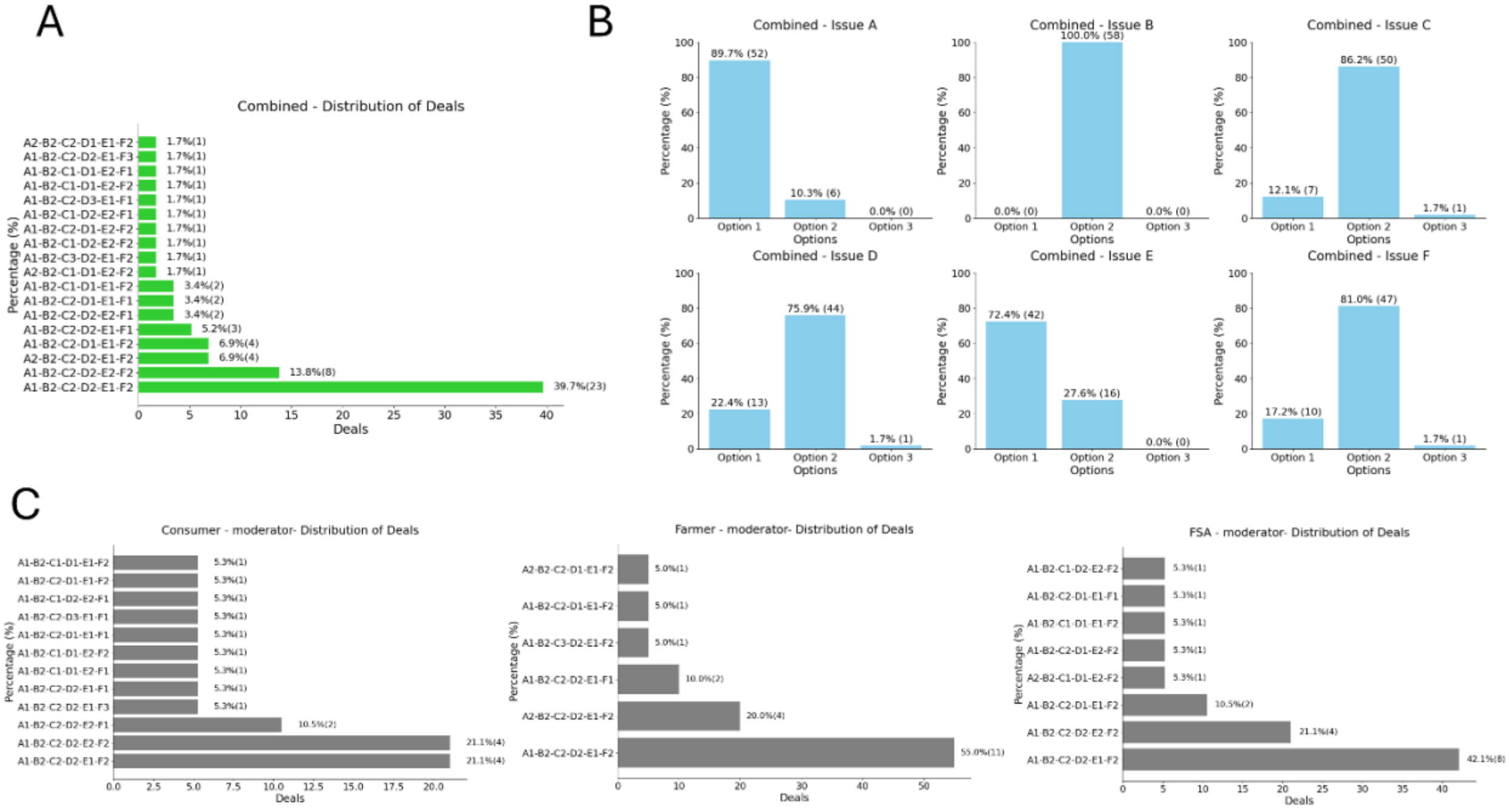

A: Distribution of proposed deals calculated from all simulation rounds. B: Distribution of option preferences per specific issue calculated from all simulation rounds. C: Distribution of proposed deals calculated from the rounds with a specific moderator (e.g. Consumer, Farmer or Food Safety Authority/FSA). The exact formulation of each issue/option matches Supplementary Text S1.

**Fig S2: Preliminary Case scenario 1: Moderator influence on the issue-option combinations**

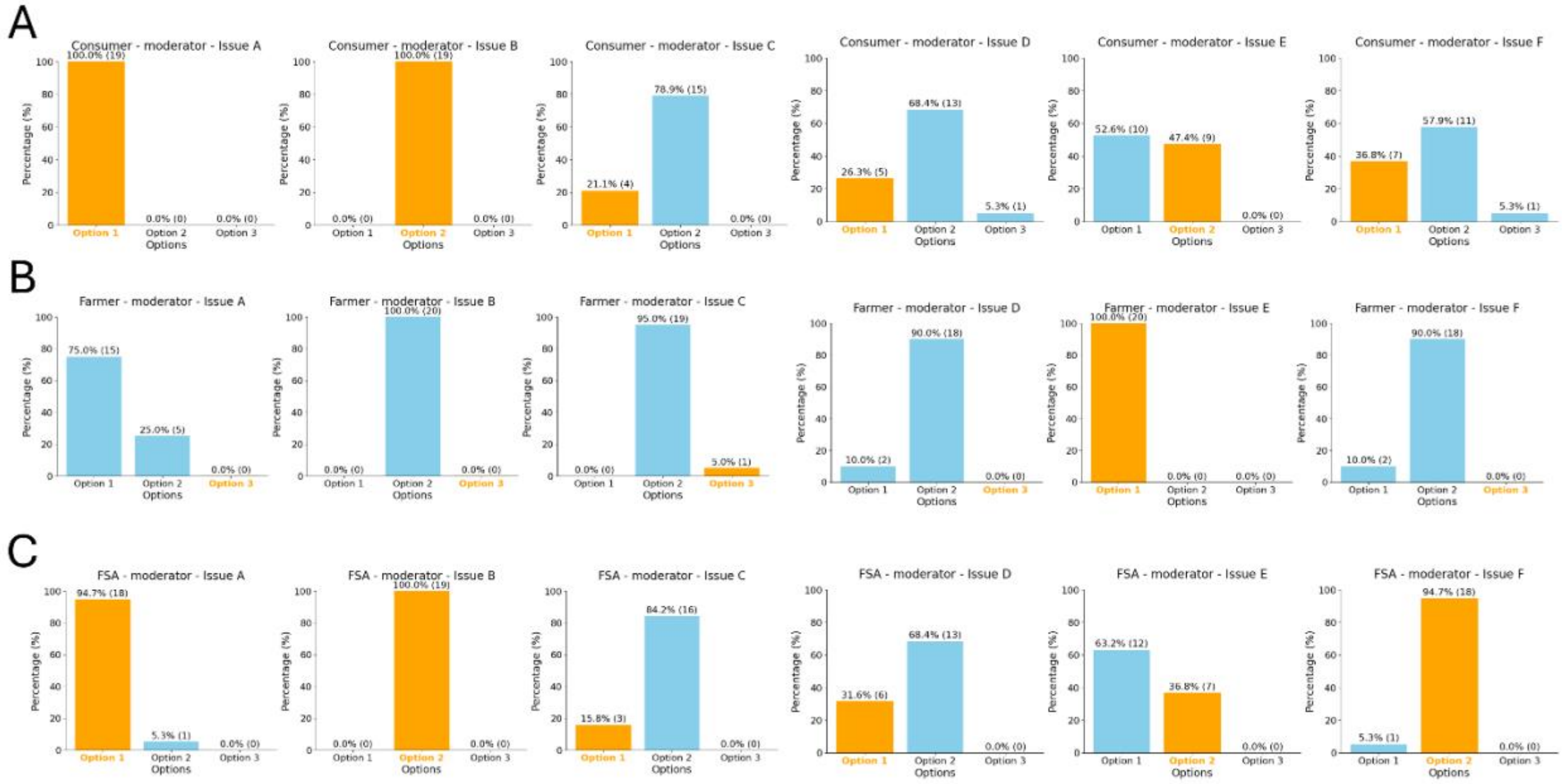

Distribution of the option preferences calculated for the simulation rounds with a moderator role assigned to (A) Consumer, (B) Farmer or (C) FSA. Orange color highlights the preferred option for the corresponding issue for a chosen moderator.

**Fig S3. Preliminary Case scenario 2: Combined results – distribution of unique issue-option combinations**

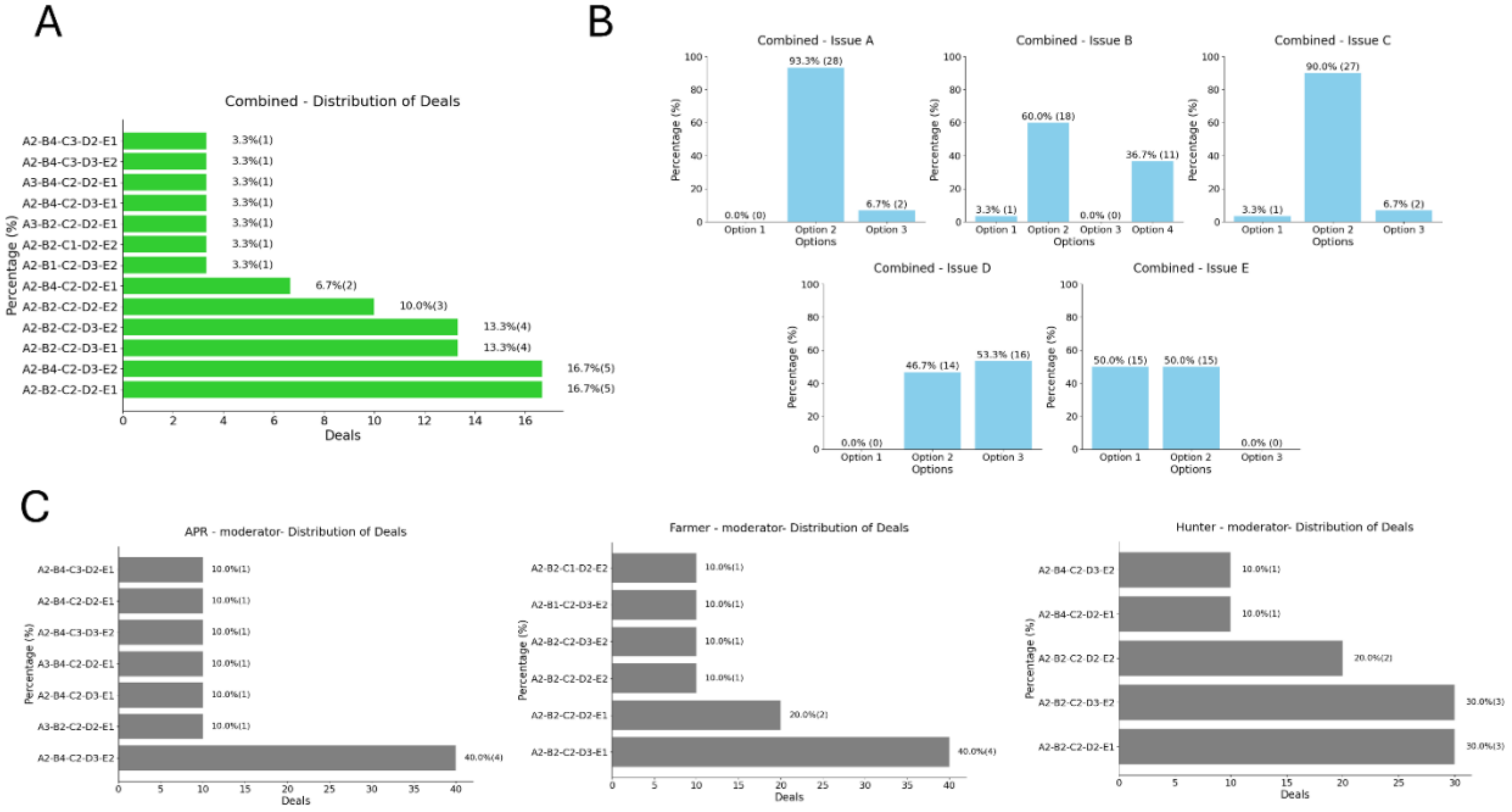

A: Distribution of proposed deals calculated from all simulation rounds. B: Distribution of option preferences per specific issue calculated from all simulation rounds. C: Distribution of proposed deals calculated from the rounds with a specific moderator (e.g. Animal Protection Representative/APR,  Farmer or Hunter,). The exact formulation of each issue/option matches Supplementary Text S2.

**Fig S4: Preliminary Case scenario 2: Moderator influence on the issue-option combinations**

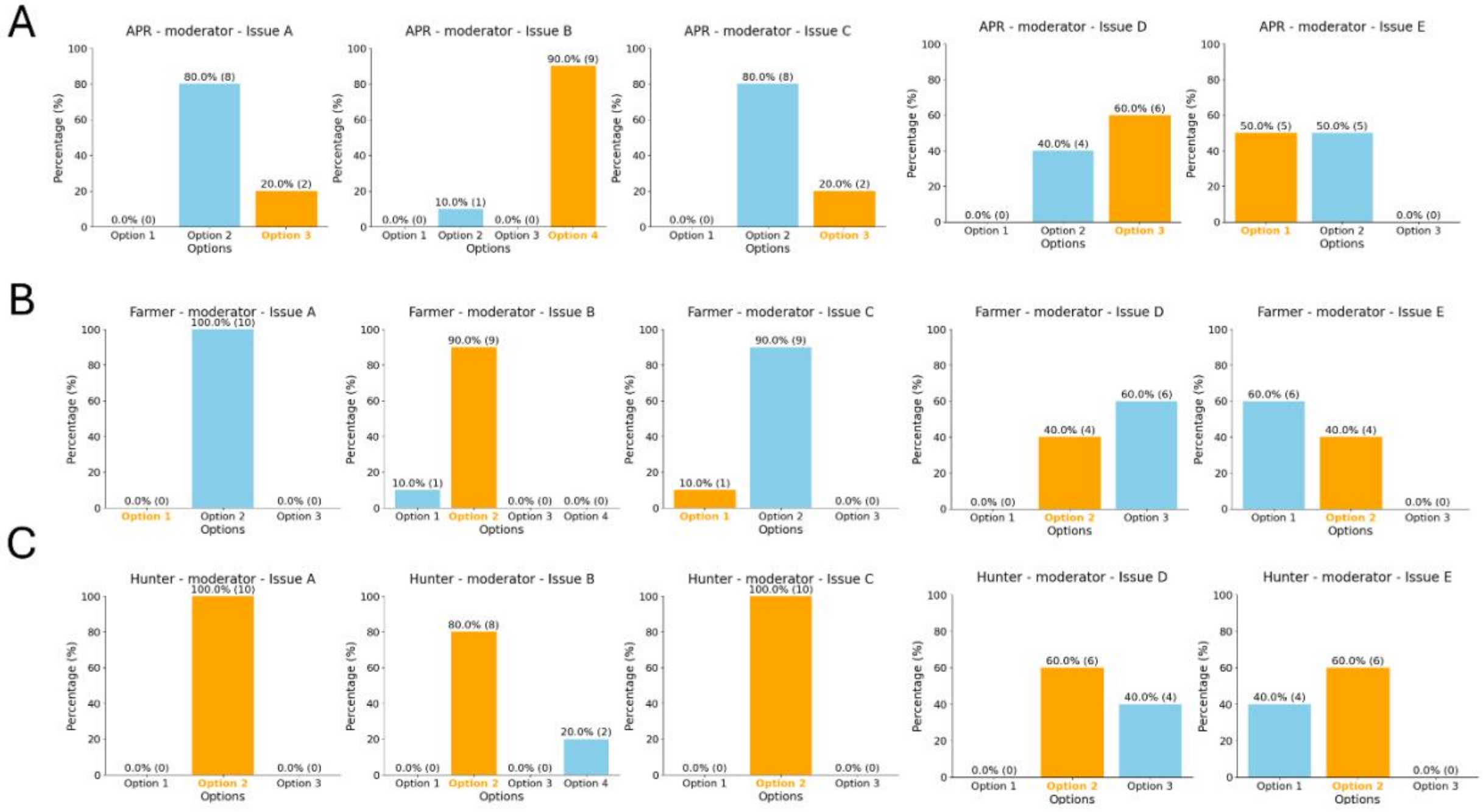

Distribution of the option preferences calculated for the simulation rounds with a moderator role assigned to (A) APR, (B) Farmer or (C) Hunter. Orange color highlights the preferred option for the corresponding issue for a chosen moderator.

**Fig S5. Practical implementation of the negotiation game post-analysis**

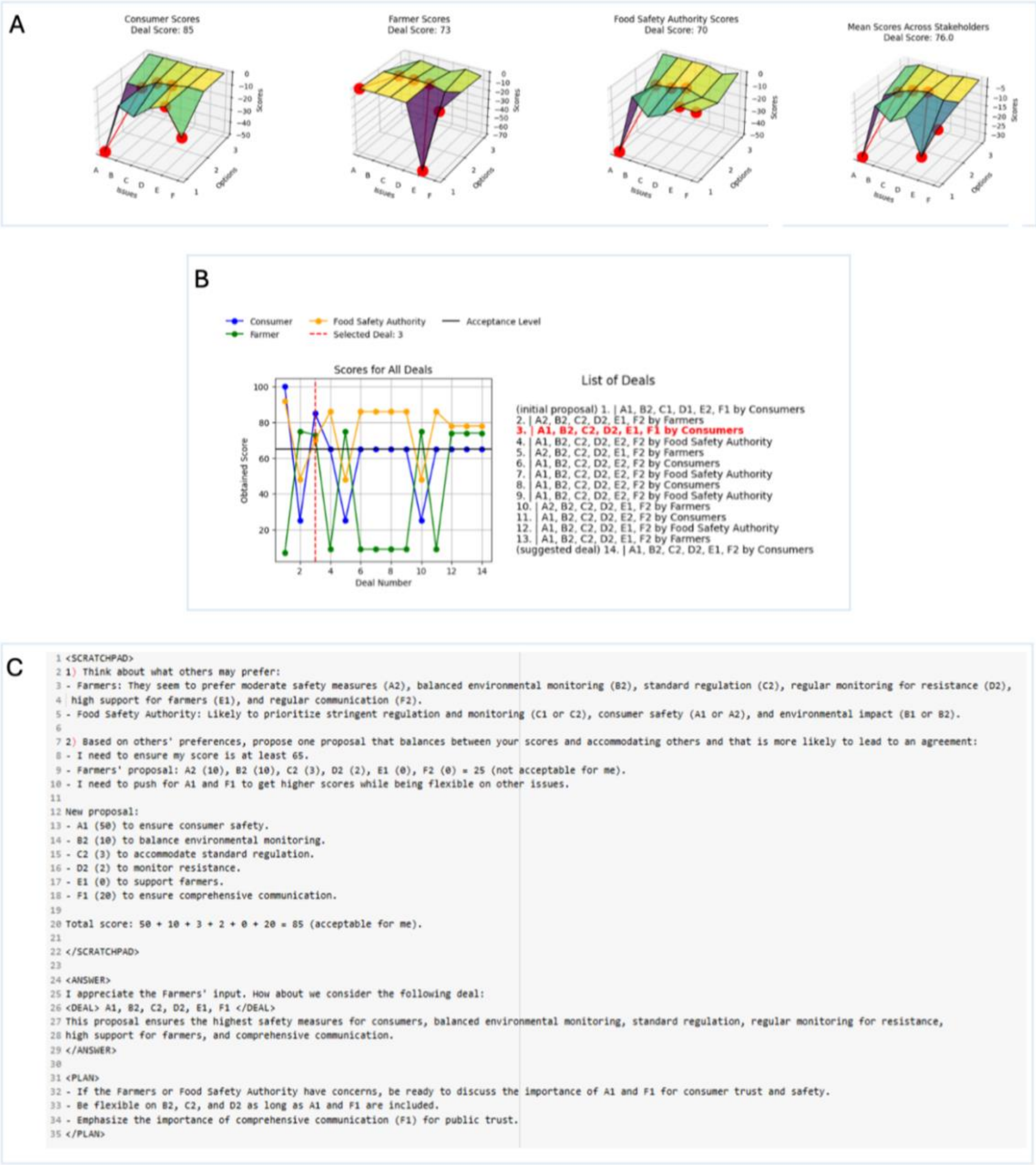

(A) Stakeholders' preferences can be represented as 3D surfaces defined by Issues, Options, and Scores (with negative values used for visualization). Each deal appears as a line on these surfaces. The surface representing the average scores across stakeholders reflects the combined interests.
(B) Each proposed deal results from 12 rounds of negotiations. In each round a stakeholder either endorses a previously proposed deal or suggests a new one.
(C) Each deal proposed in a round is linked with a scratchpad that outlines the rationale behind the proposal and plans for the next steps.

**Fig S6. October Case scenario 3 (*Campylobacter spp.*): Combined results – distribution of unique issue-option combinations**

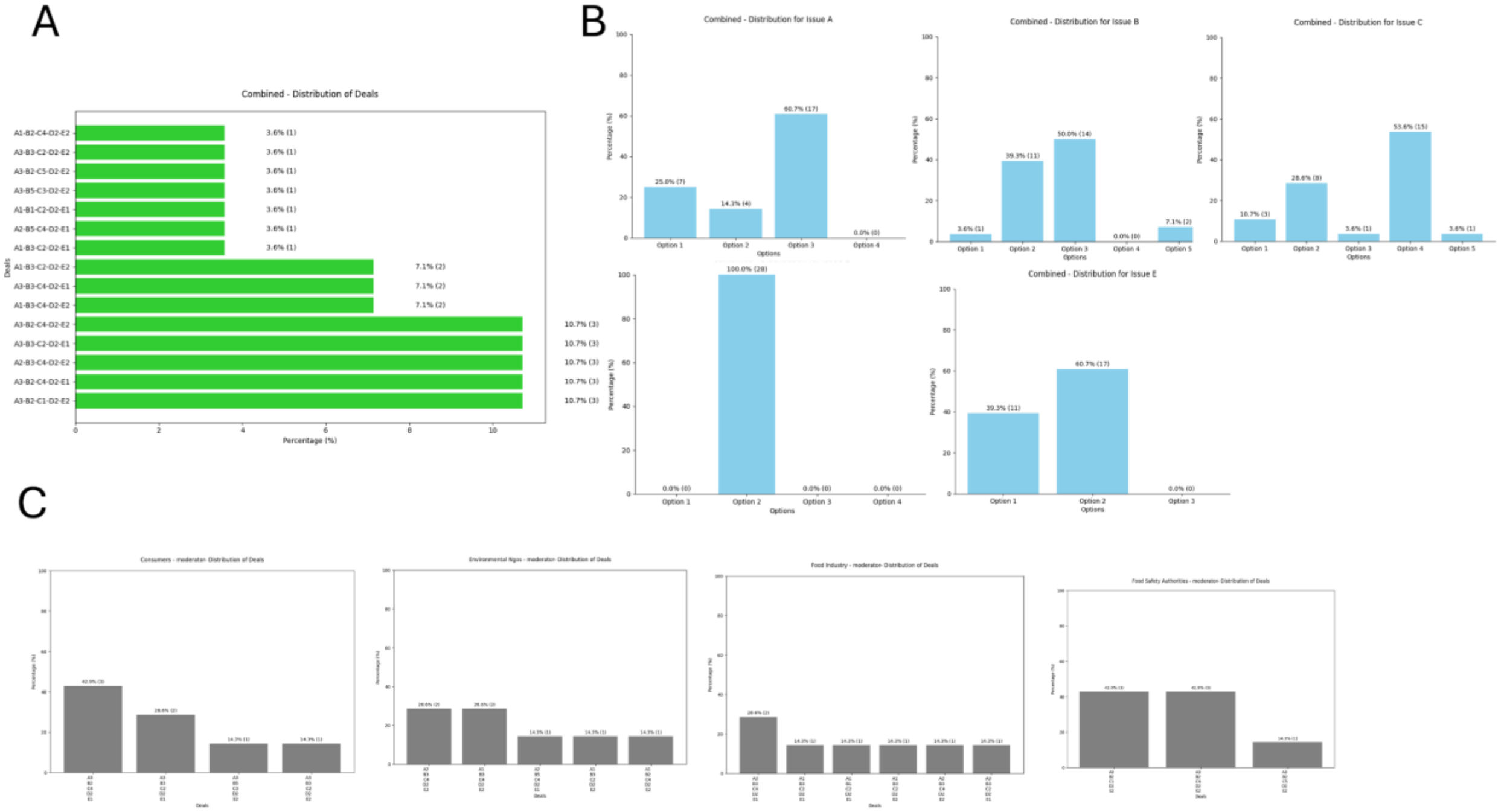

A: Distribution of proposed deals calculated from all simulation rounds. B: Distribution of option preferences per specific issue calculated from all simulation rounds. C: Distribution of proposed deals calculated from the rounds with a specific moderator

**Fig S7: October Case scenario 3 (*Campylobacter spp.*): Moderator influence on the issue-option combinations**

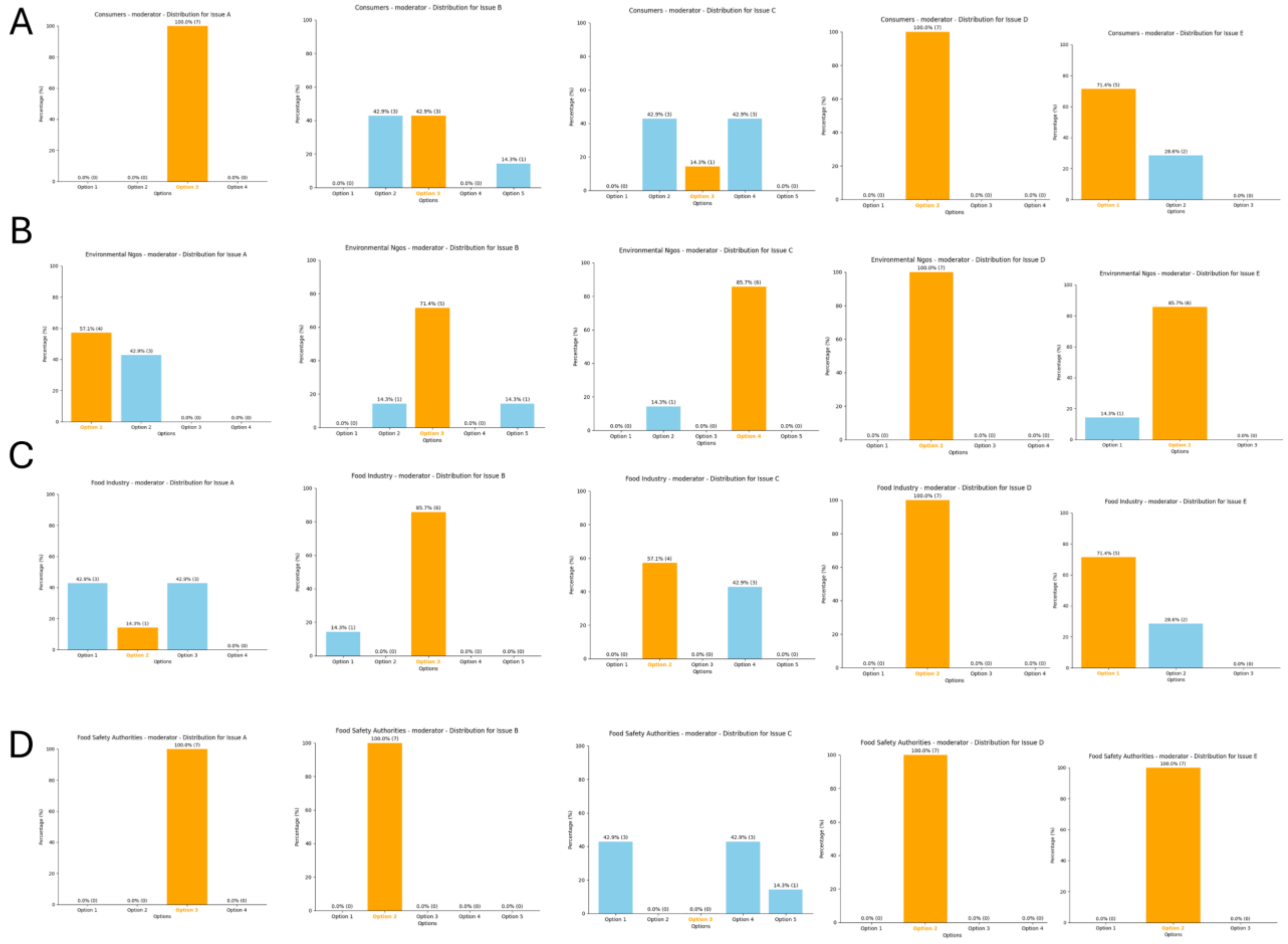

Distribution of the option preferences calculated for the simulation rounds with a moderator role assigned to each of the stakeholders. Orange color highlights the preferred option for the corresponding issue for a chosen moderator.

**Fig S8. October Case scenario 4 (OPM): Combined results – distribution of unique issue-option combinations**

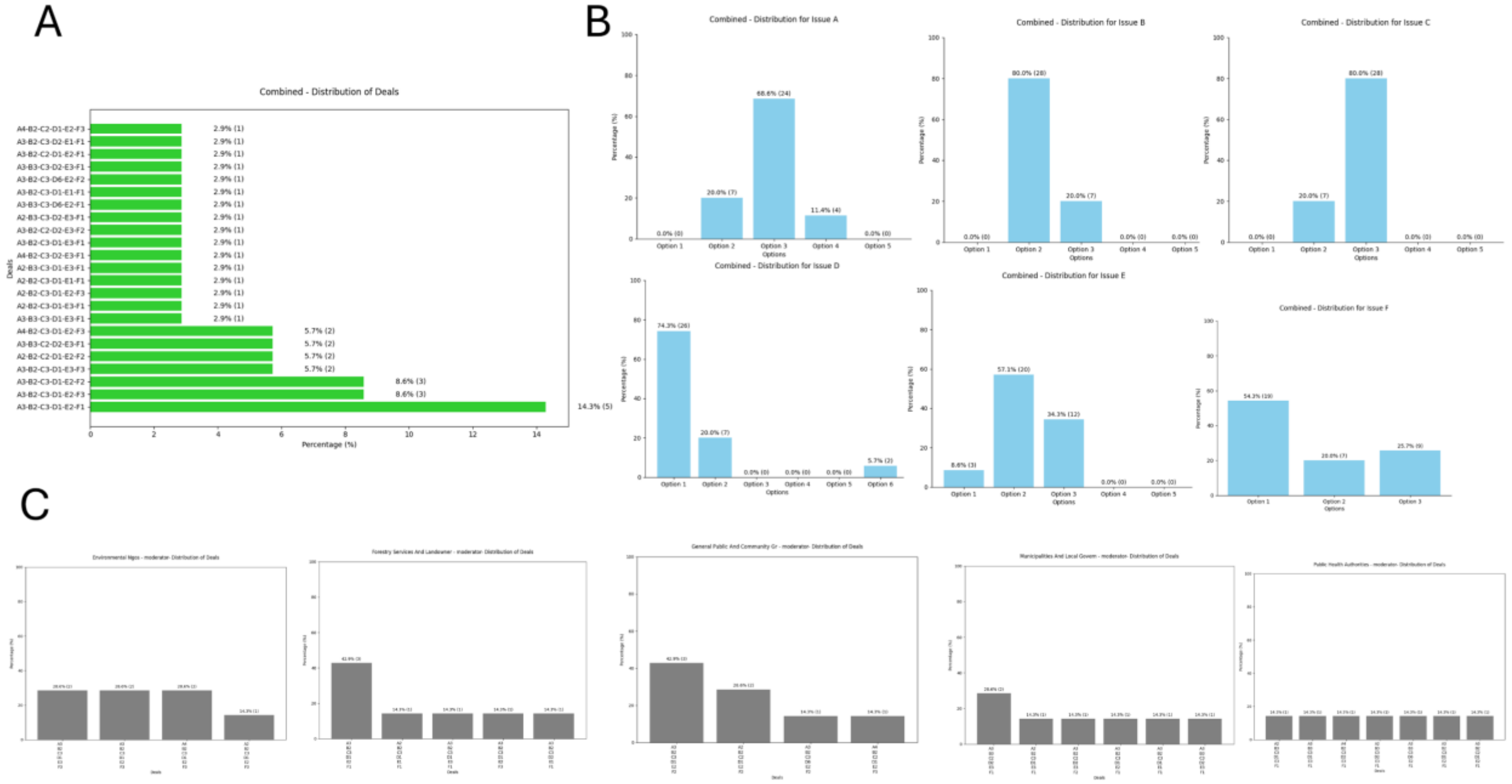

A: Distribution of proposed deals calculated from all simulation rounds. B: Distribution of option preferences per specific issue calculated from all simulation rounds. C: Distribution of proposed deals calculated from the rounds with a specific moderator

**Fig S9: Seminar Case scenario 2 (OPM): Moderator influence on the issue-option combinations**

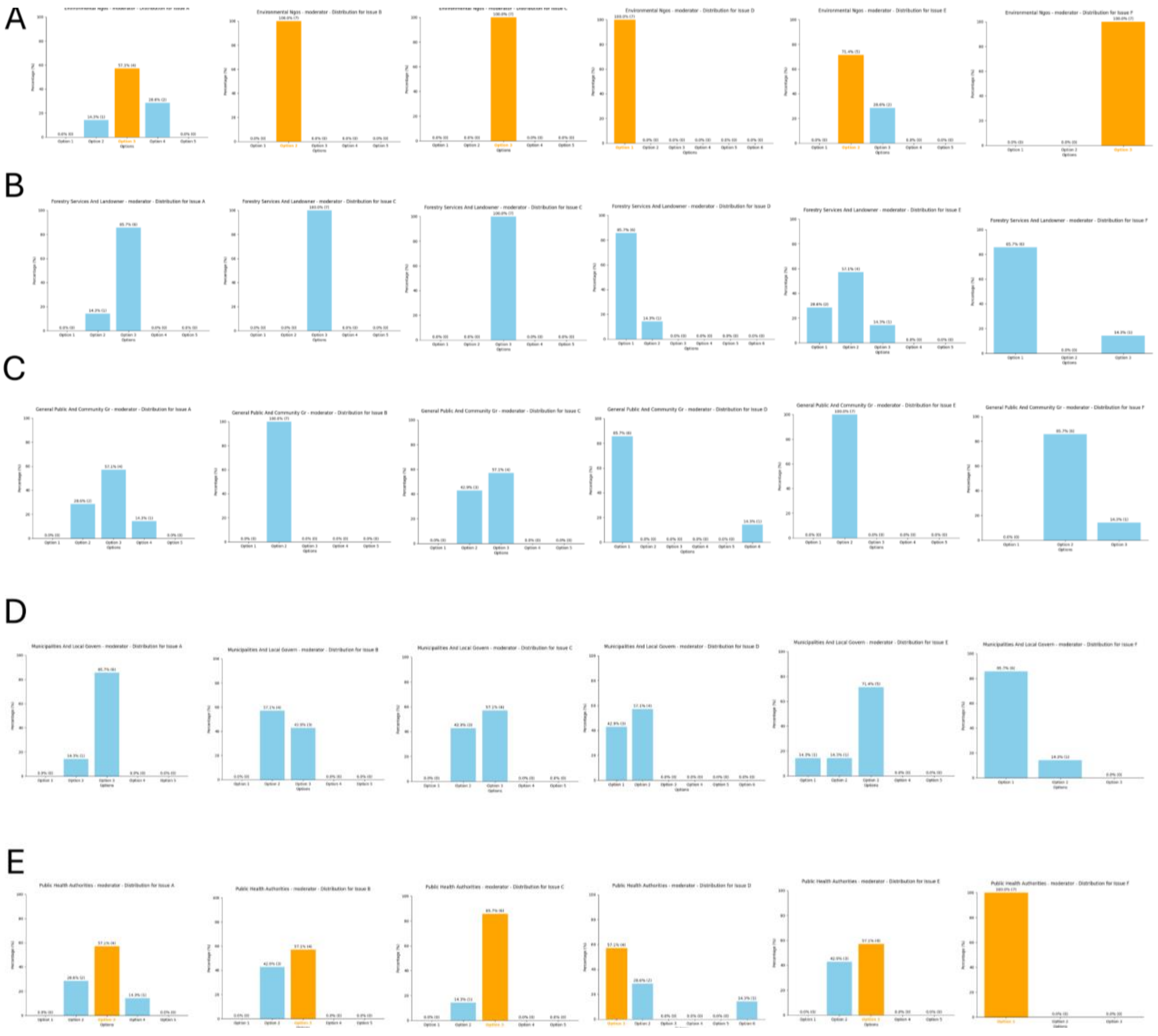

Distribution of the option preferences calculated for the simulation rounds with a moderator role assigned to each of the stakeholders. Orange color highlights the preferred option for the corresponding issue for a chosen moderator.

**Fig S10. Coalition analysis for Case scenario 2 (OPM).**

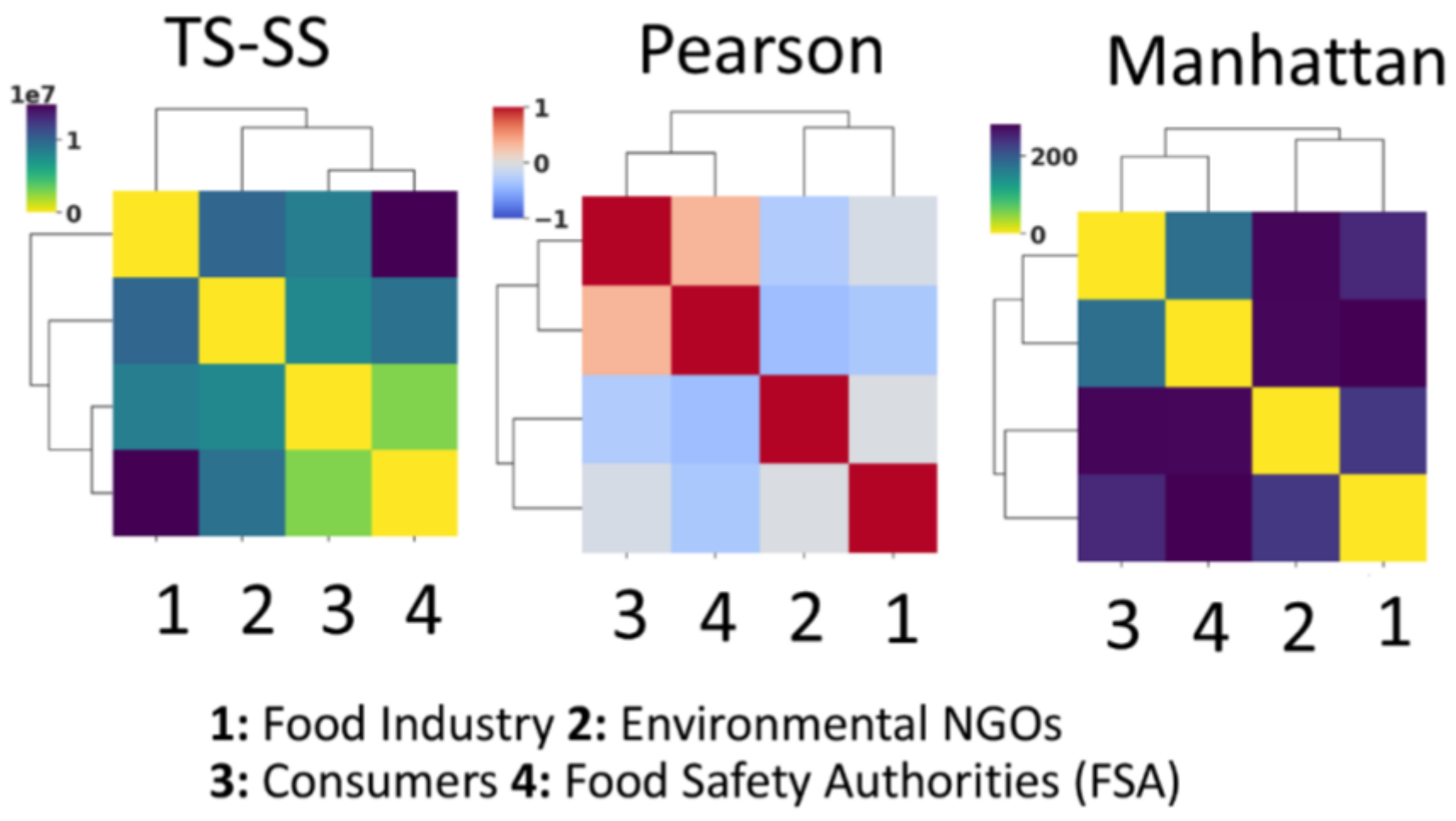

**Fig S11. Emergence of NBS in *Campylobacter spp.* case during adaptation analysis**

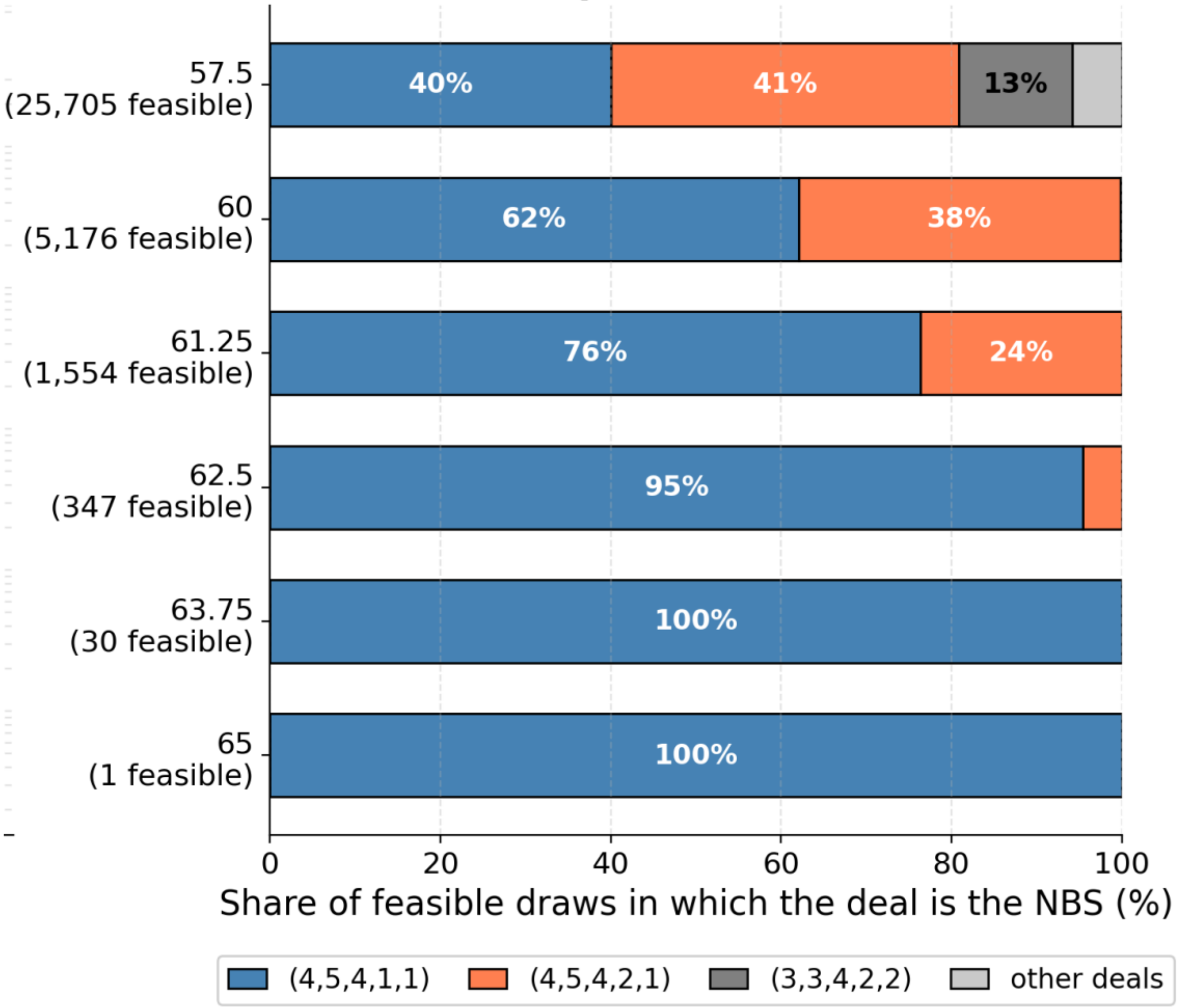

**Table S1. Sensitivity analysis for OPM case. Matched with values from Figure 4C**
The following structure is kept: value at original scores | Monte Carlo median [5th percentile−95th percentile]

| ID | Hamming distance | $L_1$ distance | $L_2$ distance | Nash loss (%) |
|---|---|---|---|---|
| **NBS** | **0 \| 2 [0–4]** | **0.0 \| 22.7 [0.0–32.8]** | **0.0 \| 12.6 [0.0–17.5]** | **0.0 \| 15.2 [0.0–74.9]** |
| R1(323121) | 2 \| 3 [1–4] | 34.0 \| 42.0 [30.8–58.9] | 20.4 \| 22.0 [17.9–28.9] | — \| 76.2 [42.8–97.7] |
| R2(323122) | 1 \| 3 [1–4] | 33.0 \| 44.6 [30.6–62.5] | 21.5 \| 24.0 [19.0–31.8] | — \| 88.1 [59.7–99.2] |
| R3(323123) | 2 \| 3 [2–4] | 44.0 \| 45.1 [31.8–70.6] | 26.1 \| 24.3 [16.6–36.3] | 95.5 \| 85.9 [58.4–99.1] |
| R4(332231) | 5 \| 4 [3–5] | 95.0 \| 100.7 [69.1–116.1] | 46.1 \| 48.9 [36.0–54.7] | — \| — |
| R5(222122) | 3 \| 4 [3–5] | 50.0 \| 61.4 [47.7–79.1] | 34.3 \| 37.9 [31.9–45.0] | — \| — |
| R6(323133) | 2 \| 3 [2–4] | 42.0 \| 41.2 [29.6–60.0] | 23.7 \| 22.0 [15.3–31.7] | 81.1 \| 81.0 [51.2–98.7] |
| R7(423123) | 3 \| 3 [2–4] | 86.0 \| 80.0 [66.6–110.6] | 46.3 \| 42.9 [35.2–58.1] | — \| — |

**Table S2. Distribution of NBS deals during sensitivity analysis**

| Deal | Count | Frequency | *H* vs baseline |
|---|---|---|---|
| (3, 2, 3, 1, 1, 2) | 2499 | 25.0% | 0 |
| (4, 2, 3, 2, 3, 2) | 1862 | 18.6% | 3 |
| (3, 2, 3, 2, 1, 3) | 1614 | 16.1% | 2 |
| (3, 2, 3, 2, 1, 2) | 1061 | 10.6% | 1 |
| (4, 2, 3, 2, 3, 1) | 908 | 9.1% | 4 |
| (3, 2, 3, 2, 1, 1) | 800 | 8.0% | 2 |
| (3, 2, 3, 1, 1, 1) | 558 | 5.6% | 1 |
| (4, 2, 3, 2, 2, 1) | 529 | 5.3% | 4 |
| (4, 2, 3, 2, 1, 2) | 60 | 0.6% | 2 |
| (3, 2, 3, 2, 2, 3) | 37 | 0.4% | 3 |
| (3, 2, 3, 2, 5, 3) | 26 | 0.3% | 3 |
| (4, 2, 3, 2, 2, 2) | 20 | 0.2% | 3 |
| (3, 2, 3, 2, 3, 3) | 13 | 0.1% | 3 |
| (4, 2, 3, 2, 5, 2) | 10 | 0.1% | 3 |
| (4, 2, 3, 2, 5, 1) | 1 | 0.0% | 4 |
| (4, 2, 3, 2, 1, 1) | 1 | 0.0% | 3 |

**Supplementary Material: Additional References – if not referred to before**